\documentclass[a4paper,11pt]{article}
\usepackage{jheppub}           

\usepackage[T1]{fontenc}       
\usepackage[utf8]{inputenc}    
\usepackage{microtype}
\usepackage{csquotes}

\usepackage{amsmath,amssymb,mathtools,bm}
\usepackage{microtype}         
\usepackage{graphicx,booktabs} 
\usepackage{xcolor}            
\usepackage{slashed}
\usepackage[nameinlink,capitalise]{cleveref} 
\usepackage{tikz}
\usepackage[compat=1.1.0]{tikz-feynhand}
\usetikzlibrary{decorations.pathmorphing}
\usetikzlibrary{arrows.meta,positioning,calc}
\usepackage{adjustbox}
\newlength{\boxw}
\tikzset{
  box/.style={
    draw, rounded corners, align=center,
    text width=\boxw, minimum height=12mm,
    inner sep=3.5pt, line width=0.8pt
  },
  arr/.style={-{Latex[length=2.6mm]}, line width=0.9pt, shorten >=1.2pt, shorten <=1.2pt}
}
\newcommand{\makeports}[1]{%
  \coordinate (#1N) at ($ (#1.north)!0.08!(#1.center) $);
  \coordinate (#1S) at ($ (#1.south)!0.08!(#1.center) $);
  \coordinate (#1E) at ($ (#1.east)!0.08!(#1.center) $);
  \coordinate (#1W) at ($ (#1.west)!0.08!(#1.center) $);
}

\DeclarePairedDelimiter{\pqty}{(}{)}     
\DeclarePairedDelimiter{\bqty}{[}{]}     
\DeclarePairedDelimiter{\Bqty}{\{}{\}}   
\DeclarePairedDelimiter{\abs}{\lvert}{\rvert}
\DeclarePairedDelimiter{\norm}{\lVert}{\rVert}

\newcommand{\dd}{\mathop{}\!\mathrm{d}}
\newcommand{\dv}[3][]{\frac{\mathrm{d}^{#1} #2}{\mathrm{d} #3^{#1}}}

\DeclareMathOperator{\tr}{tr}

\DeclareMathOperator{\Disc}{Disc}
\DeclareMathOperator{\diag}{diag}
\let\Re\relax
\DeclareMathOperator{\Re}{Re}
\let\Im\relax
\DeclareMathOperator{\Im}{Im}

\newcommand{\cO}{\mathcal{O}}

\newcommand{\YB}{Y_B}

\newcommand{\gstar}{g_{\star}}
\newcommand{\Besselk}{\mathrm{K}}

\newcommand{\CNB}{C_{NB}}      
\newcommand{\ONB}{\mathcal{O}_{NB}}      
\newcommand{\CEG}{C_{e\gamma}}
\newcommand{\OEG}{\mathcal{O}_{e\gamma}}

\newcommand{\beq}{\begin{equation}}
\newcommand{\eeq}{\end{equation}}

\numberwithin{equation}{section}

\enlargethispage{\baselineskip}

\allowdisplaybreaks

\preprint{RESCEU-23/25}

\title{Electromagnetic leptogenesis --- an EFT-consistent analysis
via Wilson coefficients. Part II. Low-scale, resonant regime}

\author[a]{Rin Takada}
\affiliation[a]{Research Center for the Early Universe (RESCEU), Graduate School of Science, The University of Tokyo, 7-3-1 Hongo, Bunkyo, Tokyo 113-0033, Japan}
\emailAdd{takada-rin@resceu.s.u-tokyo.ac.jp}

\abstract{
We study electromagnetic leptogenesis (EMLG) in the low-scale, resonant regime within a fully effective-field-theory (EFT)--consistent framework. Starting from a UV Lagrangian and performing a one-loop matching to obtain the gauge-invariant dipole operator $\mathcal{O}_{NB}=(\bar{L}\sigma^{\mu\nu}N)\tilde{H}B_{\mu\nu}$ with its Wilson coefficient $C_{NB}$, we implement the Pilaftsis--Underwood resummation for the self-energy contributions to the CP asymmetries and evolve a flavour covariant transport system across the electroweak window. In the quasi-degenerate limit $M_m-M_i\simeq\varGamma_m/2$, the CP-odd source acquires a Breit--Wigner--form enhancement whereas the washout saturates, with the freeze-out baryon asymmetry achieving $Y_B^{\rm FO}\gtrsim 10^{-10}$ across the strong-washout regime, comfortably above the observed value $Y_B^{\rm obs}\simeq 8.7\times 10^{-11}$. This establishes resonant EMLG at $\mathcal{O}(100)~{\rm GeV}$ as a viable EFT-based mechanism for the baryon asymmetry of the Universe.
}

\arxivnumber{2510.21089}

\begin{document}
\maketitle
\flushbottom

\section{Introduction}
\label{sec:1}

The observed baryon asymmetry of the Universe (BAU) calls for dynamics that satisfy Sakharov's three conditions~\cite{Sakharov:1967dj}: baryon-number violation, C and CP violation, and a departure from thermal equilibrium. In the Standard Model (SM), anomalous electroweak transitions---sphalerons~\cite{PhysRevD.28.2019, PhysRevD.30.2212}---violate $B+L$ while conserving $B-L$, thereby enabling a lepton asymmetry to be partially converted into a baryon asymmetry. Fukugita and Yanagida showed that such a lepton asymmetry can be generated by out-of-equilibrium, CP-violating decays of heavy right-handed neutrinos, thereby providing the framework of leptogenesis~\cite{FUKUGITA198645}. This framework also aligns naturally with mechanisms for generating light-neutrino masses, such as the type-I seesaw~\cite{Minkowski:1977sc, Yanagida:1979as, Gell-Mann:1979vob, PhysRevLett.44.912}. See also standard reviews~\cite{BUCHMULLER2005305, DAVIDSON2008105, FONG2012}.

Bell--Kayser--Law~\cite{PhysRevD.78.085024} first proposed the possibility of electromagnetic leptogenesis (EMLG). They showed that an electromagnetic dipole operator induces the radiative two-body decays $N\to\nu\gamma$,  $N\to\nu Z$ and thus provides the CP-violating source when loop corrections are taken into account. In the original work~\cite{PhysRevD.78.085024}, the relevant couplings are effectively treated as arbitrary inputs. In a complementary way, our previous work~\cite{Takada:2025epa} (Part~I) presented a one-loop computation, within an effective-field-theory (EFT) framework, of the Wilson coefficient $\CNB$ associated with the gauge-invariant dipole operator $\ONB=(\bar{L}\sigma^{\mu\nu}N)\tilde{H}B_{\mu\nu}$. Furthermore, we have established a systematic pipeline connecting the UV Lagrangian $\mathcal{L}_{\rm UV}$ to the observable $\YB^{\rm FO}$ at low scale, where $\YB^{\rm FO}$ denotes the freeze-out baryon asymmetry. In this EFT formulation of EMLG, the dipole operator $\ONB$ and its Wilson coefficient $\CNB$ are the central ingredients.\footnote{In this paper, we work within the EFT framework, focusing on the electroweak scale, and refer to this low-scale, resonant realisation simply as resonant EMLG; when needed, we will specify the scale explicitly (e.g.~\enquote{high-scale resonant EMLG}).}

A key feature of this framework is a common, gauge-imposed suppression: the CP-violating source $\varepsilon$ and the total widths $\varGamma$ inherit the same scaling dictated by the dipole structure. As a result, even when maximised within that setup, the freeze-out baryon asymmetry remains extremely small,
\beq
\label{eq:1.1}
Y_B^{\rm FO}\lesssim 10^{-17},
\eeq
i.e. about seven orders of magnitude below the observed value $Y_B^{\rm obs}\simeq 8.7\times 10^{-11}$~\cite{2020A&A...641A...6P}. This strong suppression arises because both the CP-violating source and the washout scale with the square of the dipole coupling $\mu^2$, so any uniform strengthening of the dipole interaction enhances production and washout in lockstep. Although the potential insufficiency of dipole-induced leptogenesis had been noted qualitatively in~\cite{Choudhury:2011gbi}, the gauge-invariant EFT treatment renders this limitation quantitative and transparent.

Motivated by the above limitation, we focus here on the quasi-degenerate regime $M_i\simeq M_m$ ($i\neq m$), where resonant enhancement from self-energy graphs becomes operative~\cite{PhysRevD.56.5431, Pilaftsis:1998pd, PILAFTSIS2004303, FLANZ1995248, FLANZ1996693, COVI1996169}. An analysis of resonant EMLG has also been presented at the TeV scale in ref.~\cite{Choudhury:2011gbi}. We implement the Pilaftsis--Underwood (PU) resummation of the internal $N_m$ propagator, which replaces the prospective singularity by a Breit--Wigner--form regulator and yields flavour-resolved loop functions $\tilde{f}_{S_a}(x)$ and $\tilde{f}_{S_b}(x)$. Working within the EFT at low scale, we include both two-body modes $N\to\nu\gamma$ and $N\to\nu Z$ consistently---an essential point because gauge invariance ties them---and we keep flavour resolution in $\varepsilon_{\alpha i}$, which avoids the cancellation of $\tilde{f}_{S_b}(x)$.

To track the out-of-equilibrium dynamics, we solve a flavour covariant transport system~\cite{SIGL1993423, DEV2014569, BHUPALDEV2015128, Dev_2015}: a $2\times 2$ matrix of densities equation for the right-handed neutrinos (retaining mixing, oscillations and decoherence) and three Boltzmann equations for the flavour asymmetries $Y_{\Delta\alpha}\coloneqq\frac{1}{3}Y_B-Y_{L_{\alpha}}$. The latter suffices in our temperature window, where charged-lepton Yukawas are in equilibrium; spectator effects are incorporated through the flavour-coupling matrix $C^{\ell(3)}$ for charged leptons.

Analytic guidance is provided by a single-parameter Breit--Wigner envelope for the self-energy enhancement,
\beq
\label{eq:1.2}
\mathcal{BW}(\Delta M)\coloneqq
\dfrac{\Delta M^2}{(\Delta M^2)^2+(M_i\varGamma_m)^2},\qquad
\Delta M^2\coloneqq M_m^2-M_i^2,
\eeq
whose maximum occurs near $\Delta M\simeq\varGamma_m/2$~\cite{PhysRevD.70.033013, DEV2017} with $\Delta M\coloneqq M_m-M_i$. The prospective singularity is regularised by the total width of the intermediate state $\varGamma_m$, while close to resonance the $\mu^2$ scaling of the source is lifted relative to the total width, allowing efficient asymmetry generation.

The analysis is kept EFT-consistent end-to-end: we start from the UV Lagrangian, perform one-loop matching to $\CNB(M_{\Psi})$, run it down to $\mu_{\rm ref}=150~{\rm GeV}$ by RGE, convert to broken-phase dipole couplings $\mu_{\alpha i}$, compute widths/projectors and resonant CP asymmetries from the same inputs, and evolve the flavour covariant transport system. This ensures that the resonant enhancement is assessed without ad-hoc regulators and that the improvement over the non-resonant EMLG is unambiguously quantified.

Finally, we emphasise that we work in a minimal \enquote{dipole-dominated} effective theory in which the gauge-invariant operator $\ONB$ provides the unique source of lepton-number violation and CP violation in the kinetic equations. At the level of the UV Lagrangian in eq.~(\ref{eq:2.1}), the SM Dirac Yukawa couplings $y_{\alpha j}$ are in general larger than the induced dipole couplings and, away from the quasi-degenerate limit, the corresponding Yukawa--induced CP asymmetries would dominate over the dipole contribution. In the resonant regime, both Yukawa-- and dipole-induced asymmetries are subject to the same width-regularised enhancement, which is bounded by $\abs{\varepsilon}\lesssim\cO(1)$ once $\abs{\Delta M}\lesssim\varGamma/2$, so the resonance cannot indefinitely amplify the Yukawa channel relative to the dipole channel. We therefore think that, for some ranges of the effective electromagnetic neutrino mass $\tilde{m}_i^{\rm EM}$ (equivalently the decay parameter $K_i$), the dipole-induced decays can account for a sizeable fraction of the BAU. In this work we consequently isolate the contribution driven by the electromagnetic dipole operator, interpret the resulting $\YB^{\rm FO}$ as the part of the BAU that can be attributed to resonant EMLG alone within this minimal setup, and leave a systematic exploration of the combined Yukawa + dipole scenario to future work.

In this work, we show the following:

(i) For a representative benchmark with $M_i=500~{\rm GeV}$ and $M_m$ chosen such that $\Delta M\simeq\varGamma_m/2$, the resonant treatment shifts the freeze-out baryon asymmetry $\YB^{\rm FO}$ upward by about nine orders of magnitude:
\beq
\label{eq:1.3}
\YB^{\rm FO}\lesssim 10^{-8}.
\eeq

(ii) In a broad interval that overlaps the neutrino-mass-motivated window ($m_2\simeq 8.6~{\rm meV}$, $m_3\simeq 50~{\rm meV}$ in the Normal Ordering (NO)), we obtain $Y_B^{\rm FO}\gtrsim Y_B^{\rm obs}$, ensuring consistency within the EFT-consistent framework.

(iii) As the effective electromagnetic neutrino mass $\tilde{m}_i^{\rm EM}$ (equivalently the decay parameter $K_i$) increases, $\YB^{\rm FO}$ exhibits a rise $\propto (\tilde{m}_i^{\rm EM})^2$ in the weak-washout regime and saturates to a plateau in the strong-washout regime.

\begin{figure}[t]
\centering
\begin{adjustbox}{max width=\linewidth}
\begin{tikzpicture}[node distance=0.035\linewidth and 0.035\linewidth]

\node[box] (uv)    {UV Lagrangian\\$\mathcal{L}_{\rm UV}$};
\node[box, right=of uv] (match) {1-loop matching\\$C_{NB}(M_\Psi)$};
\node[box, right=of match, text width=0.22\linewidth] (rge) {RGE to low-scale\\$C_{NB}(\mu_{\rm ref})$};
\node[box, right=of rge, text width=0.24\linewidth] (dip) {dipole couplings\\[0.5em]
$\displaystyle \mu_{\alpha i}=\frac{v\cos\theta_{\rm W}}{\sqrt{2}\,M_\Psi^{2}}\,C_{NB}$};

\node[box, below=0.05\linewidth of rge, xshift=0.01\linewidth] (cp) {resonant\\CP asymmetries\\
$\varepsilon_{\alpha i}^{\rm res}$\\ $f_{V_a},\; \tilde{f}_{S_a},\; \tilde{f}_{S_b}$};
\node[box, right=0.05\linewidth of cp, text width=0.24\linewidth] (wd) {widths / projectors\\
$\varGamma_i,\;K_i,\;P_\alpha$};

\node[box, below=0.035\linewidth of cp, text width=0.23\linewidth] (be) {Flavour covariant\\transport system\\(3 flavours)\\
$\rho_N,\;Y_{\Delta\alpha}$};
\node[box, below=0.03\linewidth of be,  text width=0.23\linewidth] (obs) {Observables\\$Y_B^{\rm FO}$};

\makeports{uv}\makeports{match}\makeports{rge}\makeports{dip}
\makeports{cp}\makeports{wd}\makeports{be}\makeports{obs}

\draw[arr] (uvE)    -- (matchW);
\draw[arr] (matchE) -- (rgeW);
\draw[arr] (rgeE)   -- (dipW);

\draw[arr] (dipS) -- (wdN);                  
\draw[arr] (dipS) to[out=240,in=60] (cpN);  

\draw[arr] (wdS) -- (beN);   
\draw[arr] (cpS) -- (beN);   
\draw[arr] (beS) -- (obsN);  

\end{tikzpicture}
\end{adjustbox}
\caption{Workflow of the EFT-consistent analysis adopted in this work. In the resonant analysis we evaluate CP sources with the resummed self-energy loop functions $\tilde{f}_{S_a}$, $\tilde{f}_{S_b}$, and evolve the flavour covariant transport system.}
\label{fig:1}
\end{figure}
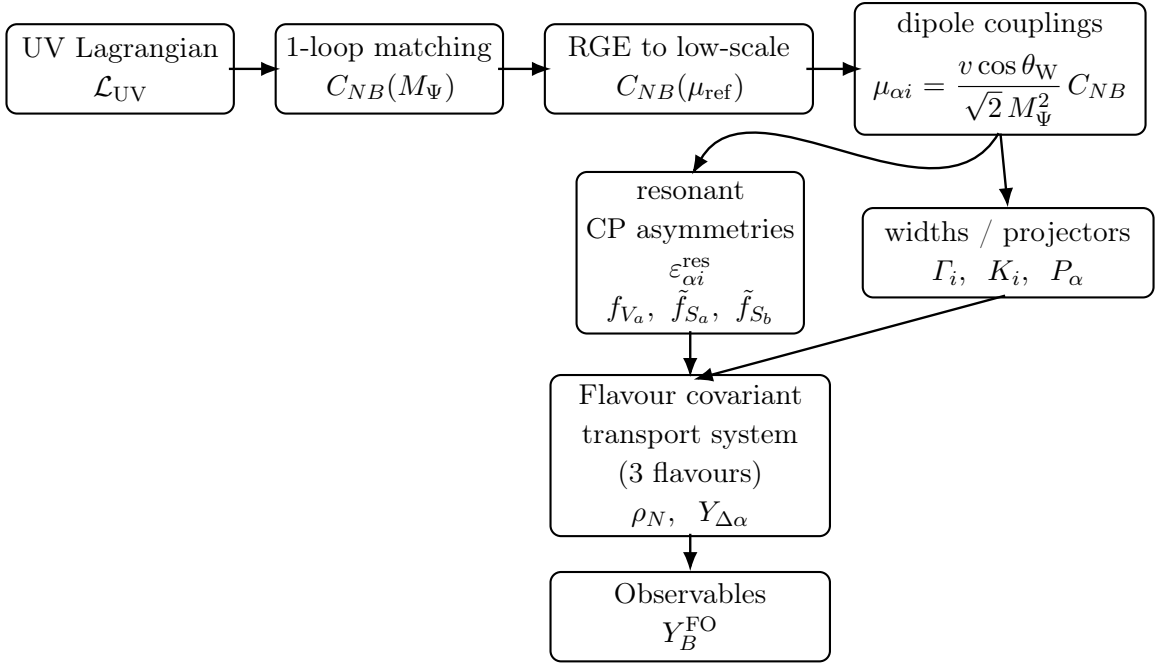

This paper is organised as follows: sec.~\ref{sec:2} sets up the EFT pipeline. In sec.~\ref{sec:3}, we derive the resummed resonant CP asymmetries, introducing the regularised loop functions and the condition for maximal enhancement. Sec.~\ref{sec:4} formulates the hybrid transport system---a flavour covariant transport equation for right-handed neutrinos coupled to three-flavour Boltzmann equations.
In sec.~\ref{sec:5}, we present numerical results: the time evolution for a given effective electromagnetic neutrino mass and a parameter scan obtained by varying it.
Sec.~\ref{sec:6} summarises the physics lessons of the EFT pipeline, explaining how resonant self-energies lift the dipole-induced suppression and reproduce the observed BAU. It also briefly describes the outlook at high scale, low-energy probes, and open theoretical improvements. Finally, in sec.~\ref{sec:7}, we briefly summarise the results and outline prospects for future work.

\section{Prerequisites}
\label{sec:2}

\subsection{UV Lagrangian}
\label{sec:2.1}

We consider a renormalisable UV completion with heavy fields whose decoupling generates gauge-invariant dimension-six operators and associated Wilson coefficients. Throughout, we work in the mass eigenbasis of the right-handed neutrinos. The Majorana mass matrix $M_N$ is diagonalised by an Autonne--Takagi decomposition with a unitary matrix, and we denote the mass eigenstates by $N_i$ ($i=1,\ldots,n_N$). Unless stated otherwise, Yukawa matrices and Wilson coefficients are expressed in this basis.

The field content extends the SM by right-handed neutrinos $N_i$, a vector-like fermion $\Psi$, and a charged scalar $S$. $D_{\mu}$ is the covariant derivative. The couplings $\lambda_i$ and $\lambda_i'$ mediate the $N_i$-$\Psi$-$S$ interactions, while $y_{\alpha j}$ is the SM Dirac Yukawa coupling for $L_{\alpha}$-$\tilde{H}$-$N_j$. We focus on the following UV Lagrangian~\cite{PhysRevD.80.013010}
\beq
\label{eq:2.1}
\begin{split}
\mathcal{L}_{\rm UV}
=\mathcal{L}_{\rm SM}
&+\bar{N}_i\mathrm{i}\slashed{\partial}N_i
+\bar{\Psi}(\mathrm{i}\slashed{D}-M_{\Psi})\Psi
+\abs{D_{\mu}S}^2-V(H,S)\\
&
-\pqty*{\lambda_i'\bar{N}_i\Psi_RS+\lambda_i\bar{\Psi}_LN_iS^{\dagger}
+\dfrac{1}{2}\bar{N}M_NN}\\
&
-\pqty*{y_{\alpha j}\bar{L}_{\alpha}\tilde{H}N_j
+y_{j\alpha}^{\ast}\bar{N}_j\tilde{H}^{\dagger}L_{\alpha}},
\end{split}
\eeq
with the scalar potential
\beq
\label{eq:2.2}
V(H,S)=M_S^2\abs{S}^2+\lambda_S\abs{S}^4
+\lambda_{HS}\abs{H}^2\abs{S}^2,
\eeq
where the three added species, $N_i$, $\Psi$, and $S$, are singlets under both $\mathrm{SU}(3)_c$ and $\mathrm{SU}(2)_L$, while their $\mathrm{U}(1)_Y$ hypercharges are 0, $+1$, and $-1$, respectively.

\subsection{Froggatt--Nielsen mechanism and naturalness}
\label{sec:2.2}

To naturally justify the tiny quasi-degeneracy required for resonance, we implement a Froggatt--Nielsen (FN) mechanism~\cite{FROGGATT1979277} in the UV completion and introduce a horizontal $\mathrm{U}(1)_{\rm FN}$ symmetry, spontaneously broken by FN fields $\Sigma$ and $\bar{\Sigma}$ carrying charges $Q_{\rm FN}(\Sigma)=+1$ and $Q_{\rm FN}(\bar{\Sigma})=-1$. Denoting vacuum expectation values by angle brackets, we define
\beq
\label{eq:2.3}
\varepsilon\coloneqq\dfrac{\langle\Sigma\rangle}{M_{\rm FN}},\qquad
\bar{\varepsilon}\coloneqq\dfrac{\langle\bar{\Sigma}\rangle}{M_{\rm FN}},
\qquad
\abs{\varepsilon},\abs{\bar{\varepsilon}}\ll 1,
\eeq
with $M_{\rm FN}$ the heavy-messenger scale. The FN selection rules then render couplings proportional to definite powers of $\varepsilon$, $\bar{\varepsilon}$, fixed by charge differences. For brevity we denote by $\varepsilon$ any small parameter of order $\abs{\varepsilon}\sim\abs{\bar{\varepsilon}}$; all scalings below are understood up to $\cO(1)$ coefficients.

We assign $\mathrm{U}(1)_{\rm FN}$ charges such that one right-handed neutrino couples at zeroth FN order whereas the other requires one FN insertion,
\beq
\label{eq:2.4}
Q_{\rm FN}(N_1)=-1,\qquad
Q_{\rm FN}(N_2)=+1,\qquad
Q_{\rm FN}(\Psi)=+1,\qquad
Q_{\rm FN}(S)=0,
\eeq
and choose lepton-doublet charges so that the SM Yukawas $y_{\alpha j}$ are unsuppressed. For the renormalisable interactions in eq.~(\ref{eq:2.1}) this implies
\beq
\label{eq:2.5}
\lambda_1\sim\cO(1),\qquad
\lambda_1'\sim\cO(1),\qquad
\lambda_2\sim\varepsilon,\qquad
\lambda_2'\sim\bar{\varepsilon},
\eeq
so that the antisymmetric combination entering the one-loop matching, $(\lambda_i\lambda_j'-\lambda_j\lambda_i')$, is $\cO(\varepsilon)$ rather than vanishing. After EWSB, the broken-phase dipole couplings eq.~(\ref{eq:2.16}) obey
\beq
\label{eq:2.6}
\mu_{\alpha 2}\propto C_{NB,\alpha 2}(\mu_{\rm ref})\propto
\cO(\varepsilon),
\qquad
\mu_{\alpha 1}\propto C_{NB,\alpha 1}(\mu_{\rm ref})\propto\cO(1),
\eeq
hence $\mu_{\alpha 2}/\mu_{\alpha 1}=\cO(\varepsilon)$. Because the mass dimension of $\mu_{\alpha i}$ equals to $-1$, the total width,
\beq
\label{eq:2.7}
\varGamma_2=\sum_{\alpha}\dfrac{\abs{\mu_{\alpha 2}}^2M_2^3}{2\pi},
\eeq
carries mass dimension $+1$ and scales as $\varGamma_2\sim\varepsilon^2M$.

For the Majorana mass matrix we obtain the generic FN texture
\beq
\label{eq:2.8}
M_N\sim M\begin{pmatrix}
\varepsilon^2&1\\
1&\bar{\varepsilon}^2
\end{pmatrix},
\eeq
whose eigenvalues are $M[1\pm\cO(\varepsilon^2)]$, so that the physical splitting is $\Delta M=M_2-M_1\sim\varepsilon^2M$. Consequently, both the mass splitting and the total width share the same $\varepsilon^2$ scaling,
\beq
\label{eq:2.9}
\Delta M\sim\varepsilon^2M,\qquad
\varGamma_2\sim\varepsilon^2M
\quad\Rightarrow\quad
\Delta M\sim\dfrac{\varGamma_2}{2},
\eeq
up to $\cO(1)$ coefficients, i.e. the resonant condition is satisfied automatically, mirroring the PU logic~\cite{PILAFTSIS2004303} while remaining fully consistent with our EFT pipeline.

The limit $\varepsilon,\bar{\varepsilon}\to 0$ restores the horizontal $\mathrm{U}(1)_{\rm FN}$ symmetry: the diagonal entries of $M_N$ vanish, $N_1$ and $N_2$ recombine into an exact Dirac pair, and the FN-suppressed dipole couplings $\mu_{\alpha 2}$ disappear. Hence, the small parameters $\varepsilon$, $\bar{\varepsilon}$ are natural in the 't Hooft sense~\cite{tHooft:1979rat}.

\subsection{Gauge-invariant operator $\ONB$ and Wilson coefficient $\CNB$}
\label{sec:2.3}

One-loop matching at $\mu=M_{\Psi}$ yields the gauge-invariant dipole operators~\cite{PhysRevD.80.013010, Patra:2011aa, Schwartz_2013}
\beq
\label{eq:2.10}
\cO_{NB,\alpha i}
=(\bar{L}_{\alpha}\sigma^{\mu\nu}P_RN_i)\tilde{H}B_{\mu\nu},\qquad
\sigma^{\mu\nu}\coloneqq
\dfrac{\mathrm{i}}{2}[\gamma^{\mu},\gamma^{\nu}],\qquad
\tilde{H}\coloneqq\mathrm{i}\sigma_2H^{\ast},
\eeq
with $B_{\mu\nu}=\partial_{\mu}B_{\nu}-\partial_{\nu}B_{\mu}$, and the Wilson coefficients
\beq
\label{eq:2.11}
C_{NB,\alpha i}(M_{\Psi})
=\sum_{j}\dfrac{g_1Y_{\Psi}(\lambda_i\lambda_j'-\lambda_j\lambda_i')y_{\alpha j}M_{\Psi}}{16\pi^2M_j}\,
\dfrac{1-r+r\ln r}{(1-r)^2},\qquad
r\coloneqq\dfrac{M_S^2}{M_{\Psi}^2}.
\eeq
The flavour structure of $\CNB$ is antisymmetric in the right-handed-neutrino indices, $C_{NB,\alpha i}\propto
(\lambda_i\lambda_j'-\lambda_j\lambda_i')y_{\alpha j}$, so only off-diagonal pairs with $i\neq j$ contribute. In particular, for a single right-handed neutrino generation, the coefficient vanishes identically; hence, at least two generations of right-handed neutrinos are required for this mechanism to operate.

\subsection{$\Delta L=0$ four-point effective operators}
\label{sec:2.4}

Beyond the gauge-invariant dipole $\ONB$, our one-loop matching of the UV Lagrangian in eq.~(\ref{eq:2.1}) onto operators with mass dimension $d\leqslant 6$ does not generate any additional independent, non-trivial $\Delta L=0$ operators. At this order, all $\Delta L=0$ effects can be absorbed into wave-function renormalisations of $L$, $N$, and $\tilde{H}$, shifts of the renormalisable mass and Yukawa parameters, and radiative corrections to $\Delta L=2$ operators such as the Weinberg operator. In particular, no contact four-fermion operator involving $N$ arises as a separate local interaction in the low-energy EFT.

After electroweak symmetry breaking, however, the broken-phase dipole induced by $\ONB$ mediates a set of $\Delta L=0$ scattering processes, obtained by combining the dipole vertex with standard QED and neutral-current vertices. Representative examples include $N_i\gamma\leftrightarrow\nu_{\alpha}\gamma$, $N_i\nu_{\beta}\leftrightarrow\nu_{\alpha}\nu_{\beta}$, and $N_if\leftrightarrow\nu_{\alpha}f$ with $f$ a charged SM fermion, all of which tend to drive the right-handed-neutrino ensemble closer to equilibrium without providing an independent CP-violating source. In the present study, following Part~I, we do not include these $2\leftrightarrow 2$ scatterings explicitly in the transport equations and instead adopt a minimal \enquote{dipole-dominated} setup in which the dipole-induced two-body decays and inverse decays $N_i\leftrightarrow\nu_{\alpha}\gamma$, $N_i\leftrightarrow\nu_{\alpha}Z$ are treated as the unique driver of $B-L$ production and washout. Within this approximation, the freeze-out baryon asymmetry $\YB^{\rm FO}$ quoted below should be regarded as optimistic upper bounds: incorporating the dipole-induced $\Delta L=0$ scatterings is expected to reduce the final asymmetry, and a systematic treatment of these processes is left for future work.

\subsection{Off-diagonal gauge mixing}
\label{sec:2.5}

One-loop mixing into $\cO_{NW}$ does not arise in the symmetric phase. The off-diagonal two-point function $\langle B_{\mu}W_{\nu}^3\rangle$ vanishes because the hypercharge-isospin sum over light fields satisfies $\sum_{\rm light}YT^3=0$, so the UV matching yields $\ONB$ only. After electroweak symmetry breaking (EWSB), $\ONB$ is simply expressed in the physical basis using $B_{\mu\nu}=F_{\mu\nu}\cos\theta_{\rm W}-Z_{\mu\nu}\sin\theta_{\rm W}$; this is merely a change of basis, not an extra mixing.

\subsection{RGE for $\CNB$ and coupled running of SM couplings}
\label{sec:2.6}

We adopt the 't Hooft--Feynman gauge and the $\overline{\rm MS}$ scheme. 
The Wilson coefficient obeys
\beq
\label{eq:2.12}
\mu\dfrac{\dd}{\dd\mu}\CNB=-2\gamma_{NB}\CNB,\qquad
\gamma_{NB}=\dfrac{1}{2}(\gamma_L+\gamma_N+\gamma_H+\gamma_B),
\eeq
where $\gamma_i$ are the field anomalous dimensions. From the one-loop two-point functions the Wilson coefficient satisfies the RGE
\beq
\label{eq:2.13}
\mu\dfrac{\dd}{\dd\mu}C_{NB,\alpha i}(\mu)
=-\dfrac{1}{16\pi^2}\pqty*{\dfrac{91}{12}g_1^2+\dfrac{9}{4}g_2^2-3y_t^2}
C_{NB,\alpha i}(\mu).
\eeq
All our numerical results jointly evolve $\{g_1,g_2,g_3,y_t\}$ together with $\CNB$ between $\mu=M_{\Psi}$ and $\mu_{\rm ref}=150~{\rm GeV}$~\cite{MACHACEK198383, MACHACEK1984221}.

\subsection{Broken-phase dipole couplings}
\label{sec:2.7}

After EWSB, the gauge-invariant operator $\ONB$ is expressed in the physical basis $B_{\mu\nu}=F_{\mu\nu}\cos\theta_{\rm W}-Z_{\mu\nu}\sin\theta_{\rm W}$ with $\tilde{H}\to(v/\sqrt{2}\;\;0)^{\intercal}$. We therefore work with a single broken-phase dipole $\ONB$ that simultaneously induces $N_i\to\nu_{\alpha}\gamma$ and $N_i\to\nu_{\alpha}Z$. 

The effective interaction then reads~\cite{Grzadkowski:2010es, BRIVIO20191}
\beq
\label{eq:2.14}
-\mathcal{L}_{\rm EFT}\supset
\dfrac{v}{\sqrt{2}\,M_{\Psi}^2}\,C_{NB,\alpha i}(\mu_{\rm ref})
(\bar{\nu}_{\alpha}\sigma^{\mu\nu}P_RN_i)
(F_{\mu\nu}\cos\theta_{\rm W}-Z_{\mu\nu}\sin\theta_{\rm W})
+\text{h.c.},
\eeq
from which it is convenient to define the dipole coupling $\mu_{\alpha i}$ by factoring out the electroweak angles:
\beq
\label{eq:2.15}
-\mathcal{L}_{\rm EFT}\supset
\mu_{\alpha i}(\bar{\nu}_{\alpha}\sigma^{\mu\nu}P_RN_i)
(F_{\mu\nu}-Z_{\mu\nu}\tan\theta_{\rm W})
+\text{h.c.},
\eeq
with
\beq
\label{eq:2.16}
\mu_{\alpha i}\coloneqq\dfrac{v\cos\theta_{\rm W}}{\sqrt{2}\,M_{\Psi}^2}\,
C_{NB,\alpha i}(\mu_{\rm ref}).
\eeq
In the numerical calculation, the Wilson coefficient is evolved down to an electroweak reference scale. Choosing $\mu_{\rm ref}=150~{\rm GeV}$ is convenient since the gauge couplings at this scale differ from their values at $\mu=m_Z\simeq 91.2~{\rm GeV}$ by less than 1\%, well within one-loop accuracy; this choice fixes where $\CNB$ is evaluated but does not change the definition of $\mu_{\alpha i}$.

\subsection{Broken-phase rate parameters}
\label{sec:2.8}

Since the $Z$ mode is suppressed only by phase space factor and the weak-mixing angle, we include the two modes consistently in all rates and CP asymmetries.

The partial decay width into the photon mode for a given flavour reads~\cite{PhysRevD.78.085024, Law:2008yyq, PhysRevD.98.115015}
\beq
\label{eq:2.17}
\varGamma_{\gamma,\alpha i}^{\rm tree}
=\dfrac{\abs{\mu_{\alpha i}}^2M_i^3}{2\pi},
\eeq
and summing over $\alpha$ gives the total photon-mode width $\varGamma_{\gamma,i}^{\rm tree}$. The partial width into the $Z$ mode per flavour is
\beq
\label{eq:2.18}
\varGamma_{Z,\alpha i}^{\rm tree}
=\dfrac{\abs{\mu_{\alpha i}}^2M_i^3}{2\pi}
(1-r_Z)^2(1+r_Z)\tan^2\theta_{\rm W},\qquad
r_Z\coloneqq\dfrac{m_Z^2}{M_i^2},
\eeq
and $\varGamma_{Z,i}^{\rm tree}
=\sum_{\alpha}\varGamma_{Z,\alpha i}^{\rm tree}$. The total two-body decay widths are given by
\beq
\label{eq:2.19}
\varGamma_i\coloneqq
\varGamma_{\gamma,i}^{\rm tree}+\varGamma_{Z,i}^{\rm tree}.
\eeq
The decay parameters are defined as
\beq
\label{eq:2.20}
K_i\coloneqq\dfrac{\varGamma_i}{H}\bigg|_{T=M_i},\qquad
H(T)=1.66\,\sqrt{\gstar}\,\dfrac{T^2}{M_{\rm Pl}},
\eeq
where $H(T)$ is the Hubble parameter. Correspondingly, the effective electromagnetic neutrino masses $\tilde{m}_i^{\rm EM}$ and the reference values $m_{\ast}^{\rm EM}(M_i)$ are defined such that $K_i=\tilde{m}_i^{\rm EM}/m_{\ast}^{\rm EM}(M_i)$ holds, namely~\cite{Buchmuller:2002rq}
\beq
\label{eq:2.21}
\tilde{m}_i^{\rm EM}\coloneqq
v^2M_i\sum_{\alpha}\abs{\mu_{\alpha i}}^2,\qquad
m_{\ast}^{\rm EM}(M_i)\coloneqq
\dfrac{2\pi\cdot 1.66\,\sqrt{\gstar}\,v^2}{R_Z(M_i)\,M_{\rm Pl}},
\eeq
with
\beq
\label{eq:2.22}
R_Z(M_i)\coloneqq
1+(1-r_Z)^2(1+r_Z)\tan^2\theta_{\rm W},\qquad
r_Z\coloneqq\dfrac{m_Z^2}{M_i^2}.
\eeq
For our benchmark (sec.~\ref{sec:5.1}), $K_i=1$ is realised for $\tilde{m}_i^{\rm EM}=m_{\ast}^{\rm EM}(M_i)\simeq 4.15\times 10^{-4}~{\rm eV}$.

In this work, we neglect $2\to 2$ scatterings and thermal corrections to rates and CP sources. In particular, the broken-phase dipole couplings $\mu_{\alpha i}$ also induce lepton-number-conserving $2\leftrightarrow 2$ processes such as $N_i\gamma\leftrightarrow\nu_{\alpha}\gamma$, $N_i\nu_{\beta}\leftrightarrow\nu_{\alpha}\nu_{\beta}$, and $N_if\leftrightarrow\nu_{\alpha}f$ (with $f$ a charged SM fermion), which we neglect in the present analysis; these channels open additional equilibration paths for $N_i$ but do not participate in the resonant CP-odd source and are expected to reduce the final asymmetry rather than to enhance it.

\section{CP sources in the quasi-degenerate limit}
\label{sec:3}

In the quasi-degenerate regime $M_i\simeq M_m$, the right-handed-neutrino two-point function develops overlapping poles and the naive perturbative expansion breaks down~\cite{ALTHERR1994149}. Throughout this section, we therefore separate the non-resonant expression---which displays the prospective singularity---from its consistently regularised counterpart via a Dyson-resummed propagator in the sense of Pilaftsis--Underwood (PU)~\cite{PILAFTSIS2004303}. We denote Breit--Wigner--regularised loop functions by a tilde, $\tilde{f}_{S_{a,b}}$, and label the corresponding resonant CP asymmetries by a superscript \enquote{res}, that is $\varepsilon_{\alpha i}^{\rm res}$.

\subsection{Divergence of non-resonant self-energies}
\label{sec:3.1}

\begin{figure}[t]
\centering
\includegraphics[width=\linewidth]{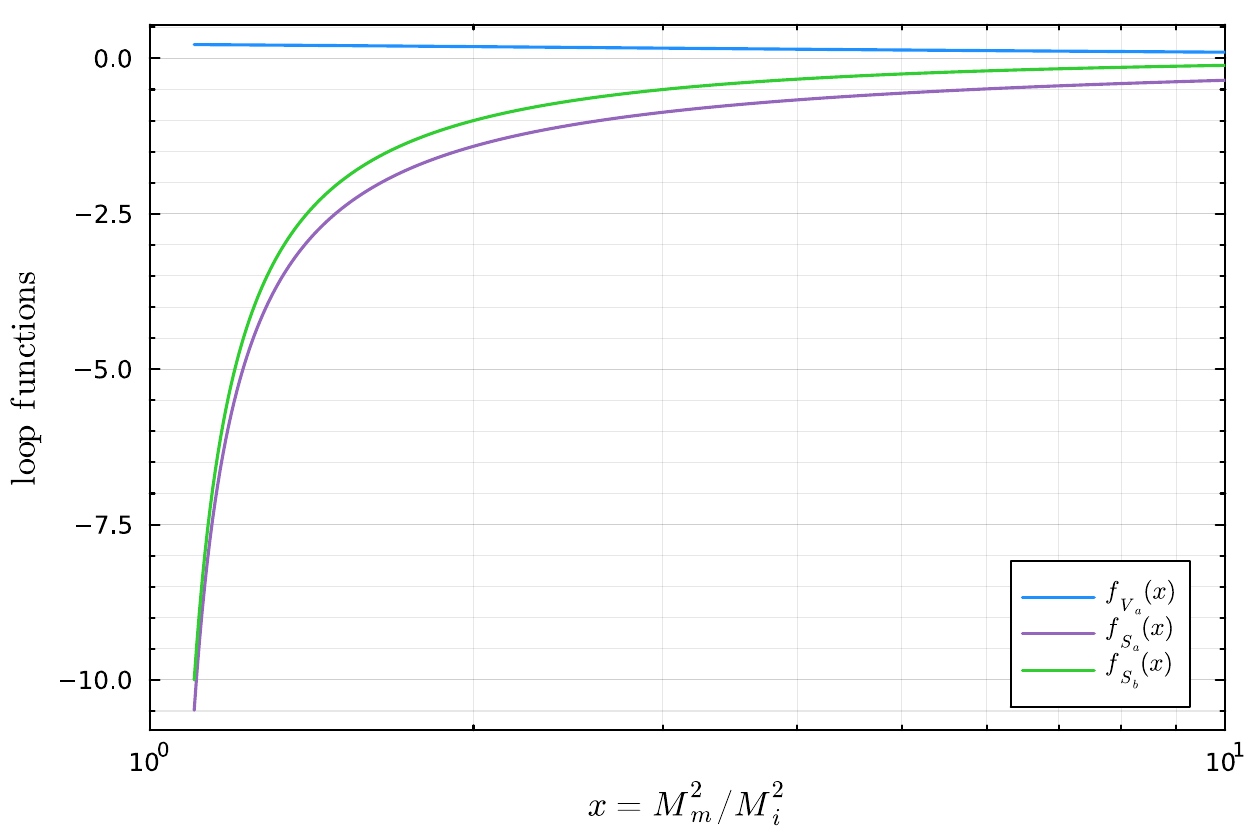}
\caption{Loop functions $f_{V_a}(x)$, $f_{S_a}(x)$, $f_{S_b}(x)$. The self-energy pieces $f_{S_a}(x)$ and $f_{S_b}(x)$ diverge as $x\to 1$, i.e. in the quasi-degenerate limit $M_i\simeq M_m$. We keep flavour resolution throughout, since under flavour summation one of the self-energy pieces, $f_{S_b}(x)$, drops out.}
\label{fig:2}
\end{figure}

The non-resonant CP asymmetries read~\cite{PhysRevD.78.085024, Law:2008yyq, Takada:2025epa}
\beq
\label{eq:3.1}
\begin{split}
\varepsilon_{\alpha i}
&=-\dfrac{M_i^2\cos^2\theta_{\rm W}}{2\pi\sum_{\beta}\abs{\mu_{\beta i}}^2}\,
R_Z(M_i)\\[0.25em]
&\qquad\times\sum_{m\neq i}
\Im\bqty*{\mu_{\alpha i}^{\ast}\mu_{\alpha m}
\Bqty*{(\mu^{\dagger}\mu)_{im}[f_{V_a}(x)+f_{S_a}(x)]
+(\mu^{\dagger}\mu)_{mi}f_{S_b}(x)}},
\end{split}
\eeq
where $x\coloneqq M_m^2/M_i^2$ and $R_Z(M_i)$ is given in eq.~(\ref{eq:2.22}). The loop functions are (see fig.~\ref{fig:2})
\beq
\label{eq:3.2}
f_{V_a}(x)=\sqrt{x}\,\Bqty*{1+2x\bqty*{1-(1+x)\ln\dfrac{1+x}{x}}},\quad
f_{S_a}(x)=\dfrac{\sqrt{x}}{1-x},\quad
f_{S_b}(x)=\dfrac{1}{1-x}.
\eeq
As $x\to 1$ the self-energy pieces $f_{S_a}(x)$ and $f_{S_b}(x)$ diverge, signalling an impending resonance~\cite{FLANZ1995248, FLANZ1996693, COVI1996169, PhysRevD.56.5431}. Flavour resolution will be crucial below: upon summing over flavours, one of the two self-energy pieces vanishes.

\subsection{Resummed propagator and regulator}
\label{sec:3.2}

\enlargethispage{\baselineskip}

In the resonant case~\cite{PILAFTSIS2004303}, the only change to the Feynman rules is the Dyson resummation of the internal $N_m$ line, which regularises the prospective pole in the self-energy graphs. Inserting the resummed propagator
\beq
\label{eq:3.3}
S_{N_m}(k)=\dfrac{\mathrm{i}(\slashed{k}+M_m)}
{k^2-M_m^2+\mathrm{i}M_m\varGamma_m},
\eeq
with the total two-body width
\beq
\label{eq:3.4}
\varGamma_m
=\sum_{\alpha}\dfrac{\abs{\mu_{\alpha m}}^2M_m^3}{2\pi}\,R_Z(M_m),
\eeq
yields the replacements~\cite{PILAFTSIS2004303, DEV2017}
\beq
\label{eq:3.5}
\begin{split}
f_{S_a}(x)&\;\longrightarrow\;
\tilde{f}_{S_a}(x)=\dfrac{\sqrt{x}\,(1-x)}{(1-x)^2+(\varGamma_m/M_i)^2},
\\[0.25em]
f_{S_b}(x)&\;\longrightarrow\;
\tilde{f}_{S_b}(x)=\dfrac{1-x}{(1-x)^2+(\varGamma_m/M_i)^2},
\end{split}
\eeq
which are directly derived in app.~\ref{sec:A}. These expressions are the actual objects used in the numerical calculations. Here, $\Delta M$ and $\Delta M^2$ are defined as
\beq
\label{eq:3.6}
\Delta M\coloneqq M_m-M_i,\qquad
\Delta M^2\coloneqq M_m^2-M_i^2
=(M_m+M_i)\Delta M
\simeq 2M_i\Delta M,
\eeq
respectively. Moreover, we also write eq.~(\ref{eq:3.5}) in terms of the physical splitting $\Delta M\coloneqq M_m-M_i$ as in
\beq
\label{eq:3.7}
\tilde{f}_{S_a}(\Delta M)
=-\dfrac{M_iM_m\Delta M^2}{(\Delta M^2)^2+(M_i\varGamma_m)^2},
\qquad
\tilde{f}_{S_b}(\Delta M)
=-\dfrac{M_i^2\Delta M^2}{(\Delta M^2)^2+(M_i\varGamma_m)^2},
\eeq
which make explicit the Breit--Wigner structure.

\subsection{Flavour-resolved asymmetries}
\label{sec:3.3}

Working flavour-resolved is essential. If flavours are summed, the $f_{S_b}$ piece vanishes by antisymmetry; with flavour separation both $\tilde{f}_{S_a}$ and $\tilde{f}_{S_b}$ survive and contribute. The flavour-resolved CP asymmetries in $N_i$ decays read
\beq
\label{eq:3.8}
\begin{split}
\varepsilon_{\alpha i}^{\rm res}
&=-\dfrac{M_i^2\cos^2\theta_{\rm W}}
{2\pi\sum_{\beta}\abs{\mu_{\beta i}}^2}\,
R_Z(M_i)\\[0.25em]
&\qquad\times
\sum_{m\neq i}\Im\Bqty*{\mu_{\alpha i}^{\ast}\mu_{\alpha m}
\bqty*{(\mu^{\dagger}\mu)_{im}[f_{V_a}(x)+\tilde{f}_{S_a}(x)]
+(\mu^{\dagger}\mu)_{mi}\tilde{f}_{S_b}(x)}}.
\end{split}
\eeq
This expression is what we employ in all scans.

\subsection{Breit--Wigner form and the condition for resonance}
\label{sec:3.4}

Eq.~(\ref{eq:3.5}) can be cast into the single-parameter envelope. Maximising the Breit--Wigner form,
\beq
\label{eq:3.9}
\mathcal{BW}(\Delta M)\coloneqq
\dfrac{\Delta M^2}{(\Delta M^2)^2+(M_i\varGamma_m)^2},
\eeq
with respect to $\Delta M^2$ yields the width-mass-splitting balance $\Delta M\simeq\varGamma_m/2$~\cite{PhysRevD.70.033013, DEV2017}.
This criterion is employed quantitatively to determine the resonant point and to define benchmark parameters.

\section{Kinetics of the flavour covariant transport equation}
\label{sec:4}

\subsection{Why the Boltzmann equation fails in the weak-washout regime}
\label{sec:4.1}

\enlargethispage{\baselineskip}

Hereafter, we specialise to $i=1$, $m=2$ and refer to $M_1$ as the clock mass in $z\coloneqq M_1/T$. In the quasi-degenerate regime $M_1\simeq M_2$, the right-handed-neutrino two-point function develops overlapping poles, and the naive perturbative expansion breaks down. A consistent resonant treatment requires a Dyson-resummed internal line embedded in a kinetic framework that can track flavour coherence and oscillations. In particular, when the decay parameter satisfies $K_1\lesssim 1$, the time scales for decays/inverse decays are comparable to the Hubble time and memory effects become relevant; the classical Boltzmann equation (BE) ceases to be quantitatively reliable. We therefore adopt the flavour covariant transport equation (FCTE)~\cite{SIGL1993423, DEV2014569, BHUPALDEV2015128, Dev_2015} in which the right-handed neutrino ensemble is evolved as a $2\times 2$ matrix of densities $\rho_N$~\cite{SIGL1993423, DEV2014569} while the charged-lepton sector, whose Yukawa interactions are in equilibrium around $T=\cO(100)~{\rm GeV}$, is treated fully flavoured and diagonal.

\subsection{Flavour covariant transport equation}
\label{sec:4.2}

Throughout this section we adopt the single-clock approximation $M_1\simeq M_2$. We take the thermal initial condition for $\rho_N$ at $z_{\rm EW}=M_1/T_{\rm EW}$ and the zero initial conditions for $Y_{\Delta\alpha}$.

Our kinetic system comprises two coupled sectors: (i) a flavour covariant transport equation for the $2\times 2$ matrix of densities associated with the right-handed neutrinos $\rho_N$ in the $\{N_1,N_2\}$ subspace, which captures coherent mixing and oscillations; and (ii) a set of three flavour-resolved Boltzmann equations for the flavour asymmetries $Y_{\Delta\alpha}=\frac{1}{3}Y_B-Y_{L_{\alpha}}$ ($\alpha=e,\mu,\tau$), which track the CP-odd production and the washout in the charged-lepton sector~\cite{SIGL1993423, DEV2014569, BHUPALDEV2015128, Dev_2015}:
\begin{subequations}
\label{eq:4.1}
\begin{align}
\dv{\rho_N}{z}
&=-\mathrm{i}\bigl[\Omega,\rho_N\bigr]
-\dfrac{1}{2}\bigl\{\mathcal{D}(z),\rho_N-\rho_N^{\rm eq}\bigr\},
\label{eq:4.1a}\\[0.25em]
\dv{Y_{\Delta\alpha}}{z}
&=\tr\bigl[\varepsilon_{\alpha}\mathcal{D}(z)(Y_N-Y_N^{\rm eq})\bigr]
-\sum_{\beta}W_{\alpha\beta}(z)Y_{\Delta\beta}.
\label{eq:4.1b}
\end{align}
\end{subequations}
Here the effective Hamiltonian,
\beq
\label{eq:4.2}
\Omega=\dfrac{z}{H(M_1)}\,
\dfrac{\diag(M_1^2,M_2^2)}{2\langle E\rangle},
\eeq
governs coherent oscillations, while $\langle E\rangle\simeq (M_1+M_2)/2$ is the averaged energy. The matrix of equilibrium densities is
\beq
\label{eq:4.3}
\rho_N^{\rm eq}=\diag(Y_{N_1}^{\rm eq},Y_{N_2}^{\rm eq}),\qquad
Y_{N_{1,2}}^{\rm eq}(z)=\dfrac{45}{4\pi^4\gstar}\,z^2\Besselk_2(z),
\eeq
and the decay matrix is given by
\beq
\label{eq:4.4}
\mathcal{D}(z)=\diag(D_1(z),D_2(z)),\qquad
D_{1,2}(z)=K_{1,2}\,z\,\dfrac{\Besselk_1(z)}{\Besselk_2(z)}.
\eeq
Flavour projectors coincide with branching fractions
\beq
\label{eq:4.5}
P_{\alpha}\coloneqq
\dfrac{\varGamma_{\gamma,\alpha i}+\varGamma_{Z,\alpha i}}{\varGamma_i}
=\dfrac{\abs{\mu_{\alpha i}}^2}{\sum_{\beta}\abs{\mu_{\beta i}}^2}.
\eeq
The washout reads~\cite{Nardi_2006, Abada_2006}
\beq
\label{eq:4.6}
W_{\alpha\beta}(z)=P_{\alpha}W_{\rm ID}(z)C^{\ell(3)}_{\alpha\beta},
\eeq
with
\beq
\label{eq:4.7}
W_{\rm ID}(z)=\dfrac{1}{4}K_1z^3\Besselk_1(z),
\eeq
and~\cite{ANTUSCH2012180}
\beq
\label{eq:4.8}
C^{\ell(3)}=\begin{pmatrix}
151/179&-20/179&-20/179\\-25/358&344/537&-14/537\\
-25/358&-14/537&344/537
\end{pmatrix}.
\eeq
Here, the single-clock approximation $M_1\simeq M_2$ is employed in the kinetic factors, namely $\rho_N^{\rm eq}$, $\mathcal{D}(z)$, and $W_{\rm ID}(z)$. This introduces only a relative error of order $\cO(\abs{\Delta M}/M_1)$. 

For orientation, it is useful to contrast the two washout regimes. In the strong-washout limit ($K_1\gg 1$), the off-diagonal coherence of the right-handed-neutrino matrix of densities is exponentially damped and the system asymptotically reduces to the classical BE of Part~I, with the CP-odd source already regularised by the Breit--Wigner form. In the weak-washout regime ($K_1\lesssim 1$), by contrast, coherent mixing and oscillations compete with damping; the matrix-of-densities treatment is then mandatory to capture the resonant enhancement quantitatively.

\subsection{On the physical interpretation of the kinetic equations}
\label{sec:4.3}

\qquad
(i) The Hamiltonian term in eq.~(\ref{eq:4.1a}), $-\mathrm{i}[\Omega,\rho_N]$, represents coherent mixing and oscillations: the phase evolves at frequency $\frac{z}{H(M_1)}\cdot\frac{\Delta M^2}{2\langle E\rangle}$, generating the off-diagonal coherence $\rho_{12}$ between $N_1$ and $N_2$~\cite{SIGL1993423, DEV2014569, BHUPALDEV2015128}.

(ii) The dissipator in eq.~(\ref{eq:4.1a}), $-\frac{1}{2}\{\mathcal{D},\rho_N-\rho_N^{\rm eq}\}$, represents dissipation and approach to equilibrium: decays and inverse decays damp coherence and drive populations towards $Y_{N_{1,2}}^{\rm eq}$ at rates set by $K_{1,2}$~\cite{SIGL1993423, BENEKE20101, DEV2014569}.

(iii) The CP-violating source in eq.~(\ref{eq:4.1b}), $\tr[\varepsilon_{\alpha}\mathcal{D}(Y_N-Y_N^{\rm eq})]$: CP-odd production tracks how much the right-handed neutrino ensemble departs from equilibrium; the resummed functions $\tilde{f}_{S_{a,b}}$ incorporate the Breit--Wigner enhancement near degeneracy~\cite{PILAFTSIS2004303, DEV2014569, BHUPALDEV2015128, Dev_2015}.

(iv) The washout matrix in eq.~(\ref{eq:4.1b}), $-\sum_{\beta}W_{\alpha\beta}Y_{\Delta\beta}$: inverse decays and spectator effects deplete each flavour asymmetry with weights $P_{\alpha}$ and the flavour-coupling matrix $C^{\ell(3)}$~\cite{Nardi_2006, Abada_2006}.

Throughout this work we follow the semi-classical flavour covariant transport framework developed in refs.~\cite{SIGL1993423, DEV2014569, BHUPALDEV2015128, Dev_2015}, in which the $2\times 2$ matrix for densities $\rho_N$ obeys a quantum kinetic equation, while the CP asymmetries are computed using the Pilaftsis--Underwood resummation of self-energy graphs~\cite{PILAFTSIS2004303}. Within this scheme the resonant Breit--Wigner structure enters only through the loop functions $\tilde{f}_{S_{a,b}}(x)$ in the source term, and we do not introduce an additional \enquote{oscillation-induced} CP source on top of the resummed $\varepsilon_{\alpha i}$. A fully consistent first-principle treatment can be formulated in the closed-time-path (CTP)/Kadanoff--Baym formalism~\cite{DEV2017}, and the precise matching between CTP-based and semi-classical descriptions is still an active topic of research. We therefore regard our implementation as one specific, widely used scheme rather than as a unique prescription, and we expect the associated theoretical uncertainty on $\YB^{\rm FO}$ to be at the level of $\cO(1)$. A dedicated CTP analysis of low-scale resonant electromagnetic leptogenesis, including a careful discussion of real-intermediate-state (RIS) subtraction~\cite{PILAFTSIS2004303} and potential double counting, is left for future work~(see sec.~\ref{sec:6.5}).

\subsection{Electroweak window and freeze-out baryon asymmetry}
\label{sec:4.4}

We model the electroweak window by the temperatures~\cite{PhysRevLett.113.141602, PhysRevD.93.025003}
\beq
\label{eq:4.9}
\begin{split}
T_{\rm EW}&\simeq 160~{\rm GeV}\quad
(\text{onset of the EWSB}),\\
T_{\rm sph}&\simeq 130~{\rm GeV}\quad
(\text{sphaleron freeze-out}),
\end{split}
\eeq
where charged-lepton Yukawa interactions are fully in equilibrium, so the three asymmetries $Y_{\Delta\alpha}$ decohere and evolve independently. Eqs.~(\ref{eq:4.1}) are evolved from $z_{\rm EW}=M_1/T_{\rm EW}$ to $z_{\rm sph}=M_1/T_{\rm sph}$. We evaluate the baryon asymmetry at sphaleron decoupling. In the fully flavoured, low-temperature regime where all charged-lepton Yukawa interactions are in equilibrium, the standard chemical-equilibrium analysis~\cite{Nardi_2006, Abada_2006} yields
\beq
\label{eq:4.10}
\YB^{\rm FO}=c_{\rm sph}\sum_{\alpha}Y_{\Delta\alpha}(z_{\rm sph}),
c_{\rm sph}=\dfrac{12}{37}.
\eeq
For comparison, in the unflavoured high-temperature regime, where flavour effects are negligible and the analysis is performed well inside the symmetric phase. Harvey and Turner~\cite{PhysRevD.42.3344} obtained the well-known relation $\YB=(28/79)Y_{B-L}$. This unflavoured result does not apply to the low-scale regime studied here, in which all SM Yukawa interactions are in equilibrium throughout the electroweak window and a fully flavoured treatment with $c_{\rm sph}=12/37$ is appropriate.

\subsection{Beyond the scope of this work}
\label{sec:4.5}

Our kinetic analysis is restricted to the electroweak-broken phase and to the two-body decays $N_i\to\nu_{\alpha}\gamma$ and $N_i\to\nu_{\alpha}Z$ induced by $\ONB$. In the symmetric phase, the same operator generates three-body decays $N_i\to\ell_{\alpha}\phi V$ with $V=B,W$~\cite{PhysRevD.78.085024}, as well as $2\leftrightarrow 2$ scatterings such as $N_i\phi\leftrightarrow\ell_{\alpha}V$, obtained by crossing these amplitudes. These channels are suppressed with respect to the broken-phase two-body modes by a combination of the extra three-body phase space, additional gauge couplings associated with the final-state gauge boson, and the fact that the effective dipole couplings are proportional to the Higgs VEV $v(T)$, which is small well above the crossover. For our benchmark choice $M_1=500~{\rm GeV}$, the lepton and baryon asymmetries are generated gradually across the interval $T\simeq 160\text{--}130~{\rm GeV}$, from the onset of the crossover down to sphaleron freeze-out, where the broken-phase description employed in this work should remain qualitatively reliable. A quantitative assessment of the symmetric-phase contribution and of the full set of $1\to 3$ decays and $2\to 2$ scatterings would require extending the EFT pipeline to higher temperatures and is beyond the scope of this work; we return to this point in sec.~\ref{sec:6.5}.

In modelling the electroweak window we adopt the standard approximation, widely used in leptogenesis studies~\cite{BUCHMULLER2005305, DAVIDSON2008105, FONG2012}, that the Higgs VEV jumps from zero to its zero-temperature value $v_0\simeq 246~{\rm GeV}$ at an effective crossover temperature and remains constant thereafter. Lattice simulations show that the electroweak transition is in fact a smooth but relatively narrow crossover~\cite{PhysRevLett.113.141602, PhysRevD.93.025003}, with $v(T)$ rising continuously from zero and reaching only a fraction of $v_0$ at $T\simeq 130~{\rm GeV}$. In our benchmark with $M_1=500~{\rm GeV}$ the relevant temperature range for baryogenesis is $T\simeq 160\text{--}130~{\rm GeV}$, so that $M_1/T\gtrsim 3$ throughout the electroweak window and $v(T)$ is already sizeable whenever the asymmetry is efficiently produced. We therefore expect that a more realistic implementation of $v(T)$, combined with the lattice-based sphaleron rate of~\cite{PhysRevLett.113.141602}, would modify the final $\YB$ only at the level of $\cO(1)$ factors rather than by orders of magnitude.

In our benchmark with $M_1=500~{\rm GeV}$, the relevant temperatures lie in the electroweak window $T\simeq 160\text{--}130~{\rm GeV}$, so that $M_1/T\sim 3\text{--}4$. In this range the right-handed neutrinos are already moderately non-relativistic, with typical thermal velocities $V\lesssim 0.7$. The difference between Fermi--Dirac and Maxwell--Boltzmann statistics in thermally averaged decay and inverse-decay rates is then at the $\cO(10\%)$ level, and we therefore work with Maxwell--Boltzmann distributions and a single temperature scale for the plasma. We expect the residual uncertainty from a more refined, momentum- and helicity-resolved quantum kinetic treatment to be of order unity and leave such an improvement for future work (see sec.~\ref{sec:6.5}).

\section{Results}
\label{sec:5}

As stated at the end of sec.~\ref{sec:2}, we neglect finite-temperature corrections to decay widths and CP asymmetries, as well as the contributions of $\Delta L=1,2$ scatterings to the washout. This matches the approximation adopted in Part~I.

\subsection{Benchmark setup}
\label{sec:5.1}

For the numerical calculations, we consider two right-handed neutrinos $N_1$, $N_2$ ($i=1$, $m=2$) in the quasi-degenerate limit. We fix $M_1=500~{\rm GeV}$, as in Part~I. For this benchmark the temperatures relevant for baryogenesis lie in the interval $T\simeq 160\text{--}130~{\rm GeV}$, so that $M_1/T\gtrsim 3$ across the electroweak window and the right-handed neutrinos are already moderately non-relativistic over the entire period during which the asymmetry is generated. The second mass $M_2$ is determined by the resonant condition (see sec.~\ref{sec:3.4}),
\beq
\label{eq:5.1}
\Delta M=M_2-M_1\simeq\varGamma_2/2,
\eeq
evaluated at $\mu_{\rm ref}=150~{\rm GeV}$ with the photon and $Z$ channels included. Equivalently, $M_2=M_1(1+\delta)$ with $\delta=\varGamma_2/(2M_1)\ll 1$. For our reference couplings, this gives $\delta=1.344\times 10^{-10}$, i.e. $M_2=(500+6.718\times 10^{-8})~{\rm GeV}$. These small quantities, $\delta$ and $\Delta M$, are associated with $\varepsilon,\bar{\varepsilon}\sim\sqrt{\delta}\sim 10^{-5}$, where both $\varepsilon$ and $\bar{\varepsilon}$ correspond to a single FN insertion, or equivalently to a charge difference $\Delta Q_{\rm FN}=1$.
In the subsequent scans, we keep $M_{1,2}$ fixed and rescale $\mu_{\alpha 1}\to\kappa\mu_{\alpha 1}$ ($\alpha=e,\mu,\tau$) while holding $\mu_{\alpha 2}$ fixed. We then recompute the width $\varGamma_1=\sum_{\alpha}\abs{\mu_{\alpha 1}}^2M_1^3/(2\pi)$ and all $\varGamma_1$-dependent quantities---notably $K_1$, $D_1(z)$, and $W_{\rm ID}(z)$---together with the PU-resummed CP sources, whereas $\varGamma_2$ and the flavour projectors $P_{\alpha}=\abs{\mu_{\alpha 1}}^2/\sum_{\beta}\abs{\mu_{\beta 1}}^2$ remain unchanged.

For heavy fields $\Psi$ and $S$, their benchmark masses are $M_S=8~{\rm TeV}$, $M_{\Psi}=10~{\rm TeV}$, the same as in Part~I. We take $\gstar=106.75$, $M_{\rm Pl}=1.22\times 10^{19}~{\rm GeV}$. Initial conditions are $Y_{\Delta\alpha}=0$ and a thermal abundance for $N_1$. Spectator effects are implemented through the $3\times 3$ flavour-coupling matrix $C^{\ell(3)}$ in eq.~(\ref{eq:4.8}), and branching fractions are collected in $P_{\alpha\beta}\simeq\diag(B_e,B_{\mu},B_{\tau})$ with $\sum_{\alpha}B_{\alpha}=1$. Baryon asymmetries are quoted as $Y_B=c_{\rm sph}\,Y_{B-L}$ with $c_{\rm sph}=12/37$.

Low-scale EMLG proceeds across the electroweak window from $z=z_{\rm EW}$ down to sphaleron freeze-out at $z=z_{\rm sph}$. All inputs are specified at $\mu_{\rm ref}=150~{\rm GeV}$, with gauge couplings and the top Yukawa evolved down to this scale. We define the freeze-out of the asymmetries to occur at sphaleron decoupling $Y_B^{\rm FO}=c_{\rm sph}\,Y_{B-L}(z_{\rm sph})$ in this work, because the $Y_{\Delta\alpha}$ curves do not form a clear plateau (cf. fig.~\ref{fig:3}).

Throughout our numerical analysis we evolve the $2\times 2$ matrix of densities $\rho_N$ for $(N_1,N_2)$, while the lepton-sector source and washout are included only for $N_1$ decays and inverse decays. The CP asymmetry is generated by the PU-resummed self-energy with an off-shell $N_2$. Direct $N_2$ contributions to the source and the washout are neglected, whereas the $N_2$ width is retained in the damping of the off-diagonal coherence via $D_{12}=\frac{1}{2}(D_1+D_2)$, in order to account for the physical dephasing of $\rho_{12}$.

\subsection{Time evolution at the fixed effective mass $\tilde{m}_1^{\rm EM}$}
\label{sec:5.2}

\begin{figure}[t]
\centering
\includegraphics[width=\linewidth]{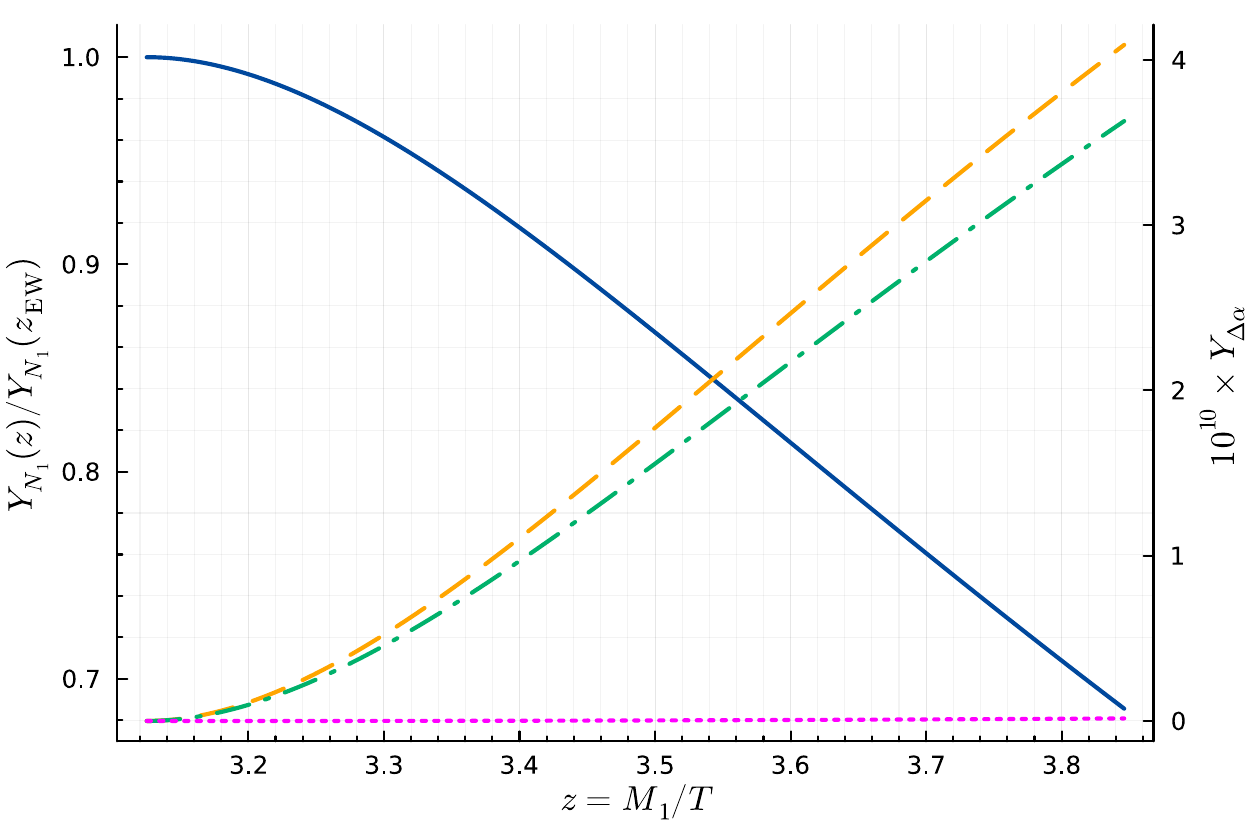}
\caption{Time evolution in the resonant regime for $\tilde{m}_1^{\rm EM}=10^{-3}~{\rm eV}$, where $\tilde{m}_1^{\rm EM}$ is defined in eq.~(\ref{eq:2.21}). The blue curve shows the depletion of $Y_{N_1}$, normalised to its value at $z=z_{\rm EW}$. The CP-odd source is computed from the PU-resummed self-energy functions $\tilde{f}_{S_{a,b}}$ at fixed dipole couplings $\mu_{\alpha 1}$. The flavour asymmetries $Y_{\Delta\alpha}$ (dashed orange for $e$, dash-dotted green for $\mu$,  dotted magenta for $\tau$) grow monotonically. The $\tau$ suppression in this benchmark originates from the small rephasing invariant~\cite{Yasaman_Farzan_2007}, $\mathcal{J}_{\tau}^{(12)}=\Im\bqty*{\mu_{\tau 1}^{\ast}\mu_{\tau 2}(\mu^{\dagger}\mu)_{12}}$.
This $\tau$ suppression is not a universal prediction and may be absent for different $\tilde{m}_1^{\rm EM}$ values.
}
\label{fig:3}
\end{figure}

We first illustrate the dynamics in the broken phase by solving the kinetic system in eq.~(\ref{eq:4.1}) from $z=z_{\rm EW}$ to $z=z_{\rm sph}$ for the benchmark $\tilde{m}_1^{\rm EM}=10^{-3}~{\rm eV}$, where $\tilde{m}_1^{\rm EM}$ is the effective electromagnetic neutrino mass defined in eq.~(\ref{eq:2.21}). Fig.~\ref{fig:3} shows the normalised abundance $Y_{N_1}(z)/Y_{N_1}(z_{\rm EW})$ (left axis) together with the flavour asymmetries $Y_{\Delta\alpha}$ (right axis; plotted as $10^{10}\times Y_{\Delta\alpha}$). The CP-odd source is computed from the resummed self-energy contributions $\tilde{f}_{S_{a,b}}$ defined in eq.~(\ref{eq:3.5}), and flavour resolution is maintained throughout. As the system departs from equilibrium after EWSB, $Y_{N_1}$ decreases monotonically, while the three $Y_{\Delta\alpha}$ build up with slopes set by $D_{1,2}(z)$ and are partially depleted by the washout matrix $W_{\alpha\beta}(z)$ (cf. sec.~\ref{sec:4.2}). By $z\simeq z_{\rm sph}$ the sum $\sum_{\alpha}Y_{\Delta\alpha}$ has reached $\cO(10^{-10})$, yielding a baryon yield $Y_B^{\rm FO}=\frac{12}{37}\sum_{\alpha}Y_{\Delta\alpha}(z_{\rm sph})$ at $\tilde{m}_1^{\rm EM}=10^{-3}~{\rm eV}$ that is of the same order as the observed value $Y_B^{\rm obs}\simeq 8.7\times 10^{-11}$~\cite{2020A&A...641A...6P}. The behaviour mirrors fig.~7 in Part~I but with a visibly larger normalisation due to the resonant enhancement captured by $\tilde{f}_{S_{a,b}}$.

\subsection{Dependence of the final baryon asymmetry on the effective mass $\tilde{m}_1^{\rm EM}$}
\label{sec:5.3}

\begin{figure}[t]
\centering
\includegraphics[width=\linewidth]{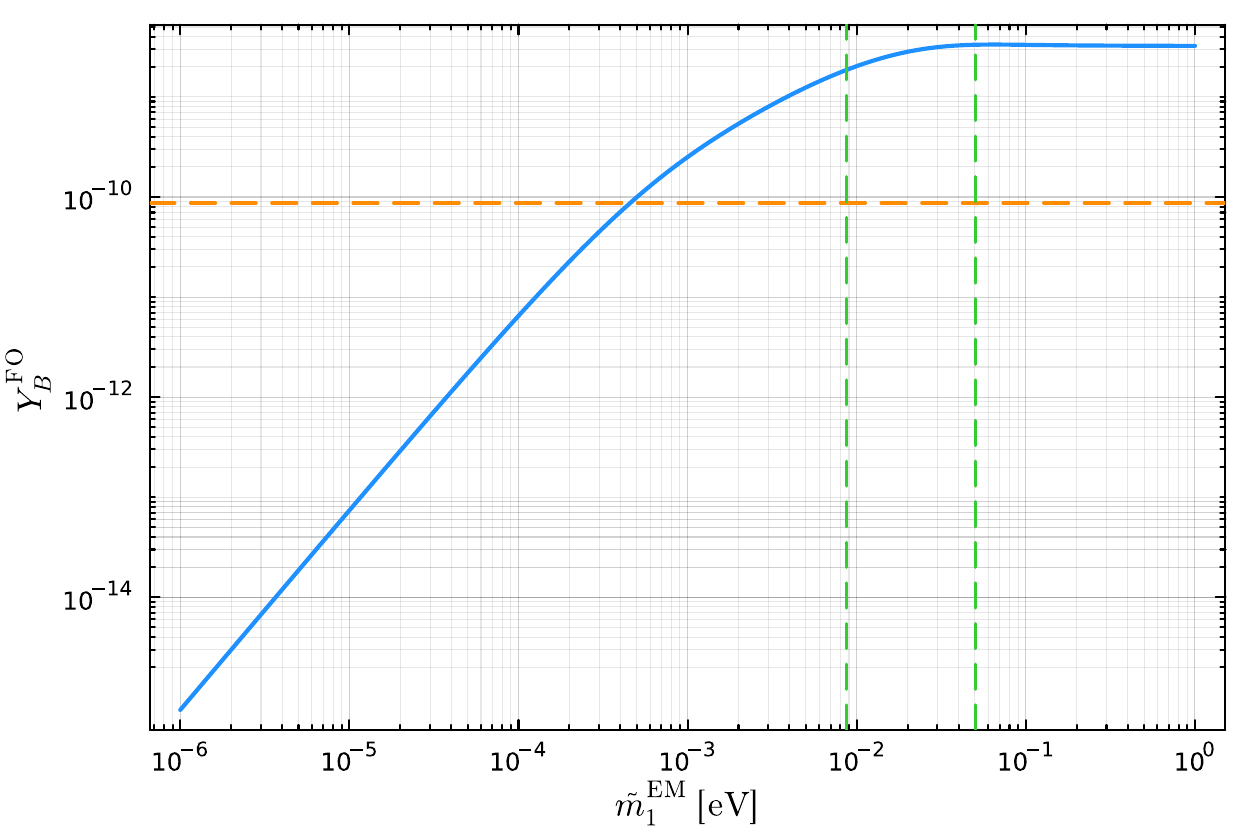}
\caption{The freeze-out baryon asymmetry $\YB^{\rm FO}$ as a function of the effective mass $\tilde{m}_1^{\rm EM}$. At the reference point $\tilde{m}_1^{\rm EM}=0.24~{\rm eV}$, we fix $M_2$ by $\Delta M\simeq\varGamma_2/2$ and thereafter vary $\tilde{m}_1^{\rm EM}$ by rescaling $\mu_{\alpha 1}$, recomputing $\varGamma_1$ and the CP sources. For the weak-washout regime $\tilde{m}_1^{\rm EM}\lesssim 4.15\times 10^{-4}~{\rm eV}$, the yield grows rapidly, $\YB^{\rm FO}\propto(\tilde{m}_1^{\rm EM})^2$, since $\varepsilon^{\rm res}\propto\mu^2$ and $D\propto K\propto\mu^2$. For the strong-washout regime $\tilde{m}_1^{\rm EM}\gtrsim 4.15\times 10^{-4}~{\rm eV}$, the CP source saturates and the washout $W(z_f)\sim 1$ drives a plateau; here we denote by $z_f$ the freeze-out of the washout. The horizontal dashed orange line marks $\YB^{\rm obs}\simeq 8.7\times 10^{-11}$~\cite{2020A&A...641A...6P}; the vertical dashed green lines show the oscillation-motivated neutrino-mass window~\cite{Esteban:2024eli}. Over the mass window we obtain $\YB^{\rm FO}\gtrsim 10^{-9}$, with a global upper envelope $\YB^{\rm FO}\lesssim 10^{-8}$ (more stringently $\YB^{\rm FO}\lesssim 4\times 10^{-9}$).
}
\label{fig:4}
\end{figure}

Here we summarise the behaviour of the freeze-out baryon yield $Y_B^{\rm FO}$ as a function of the effective electromagnetic neutrino mass $\tilde{m}_1^{\rm EM}$ when the dipole couplings $\mu_{\alpha 1}$ are rescaled at fixed $(M_1,M_2)$. We choose $M_2$ once at the benchmark such that $\Delta M\simeq\varGamma_2/2$ and then keep $M_2$ fixed while scanning; widths and CP asymmetries are recomputed with the PU-resummed expressions. The strength of washout is classified by $K_1=\varGamma_1/H(M_1)$, since the dynamics of production and erasure is governed by $N_1$.

Using eqs.~(\ref{eq:3.7})--(\ref{eq:3.8}), the flavour-resolved CP source in $N_1$ decays reads
\beq
\label{eq:5.2}
\begin{split}
&\varepsilon_{\alpha 1}^{\rm res}
=+\dfrac{M_1^2\cos^2\theta_{\rm W}}
{2\pi\sum_{\beta}\abs{\mu_{\beta 1}}^2}\,
R_Z(M_1)\\[0.25em]
&\qquad\times
\Im\bqty*{A_{\mu}^{(5)}
\dfrac{M_1M_2\Delta M^2}{(\Delta M^2)^2+(M_1\varGamma_2)^2}
+B_{\mu}^{(5)}
\dfrac{M_1^2\Delta M^2}{(\Delta M^2)^2+(M_1\varGamma_2)^2}
-A_{\mu}^{(5)}f_{V_a}\!\pqty*{\dfrac{M_2^2}{M_1^2}}},
\end{split}
\eeq
with $A_{\mu}^{(5)}
\coloneqq\mu_{\alpha 1}^{\ast}\mu_{\alpha 2}(\mu^{\dagger}\mu)_{12}$ and $B_{\mu}^{(5)}\coloneqq\mu_{\alpha 1}^{\ast}\mu_{\alpha 2}(\mu^{\dagger}\mu)_{21}$. Moreover, $f_{V_a}$ is the non-resonant vertex function in eq.~(\ref{eq:3.2}). For small $\tilde{m}_1^{\rm EM}$ (weak-washout regime $K_1\ll 1$), one has $\varGamma_2\ll\Delta M$, which is relevant to the condition of resonance. Since $\mathcal{BW}\simeq 1/\Delta M^2$ is constant, $\varepsilon_{\alpha 1}^{\rm res}\propto\mu^2\times{\rm const.}$ and the source is proportional to $\mu^2\times\mu^2\sim\mu^4$, while the effective mass $\tilde{m}_1^{\rm EM}\propto\mu^2$. Thus, $Y_B^{\rm FO}\propto(\tilde{m}_1^{\rm EM})^2$, which accounts for the nearly quadratic rise seen at the left of fig.~\ref{fig:4}.
For large $\tilde{m}_1^{\rm EM}$ (strong-washout regime $K_1\gg 1$), the envelope decreases as $\mathcal{BW}\propto 1/\varGamma_2^2\propto 1/\mu^4$ while the prefactor $\mu^2$ compensates, so the self-energy contribution to the CP source scales as $\varepsilon_{{\rm self},\alpha 1}^{\rm res}\propto 1/\mu^2$, whereas the vertex contribution grows like $\varepsilon_{{\rm vert},\alpha 1}\propto\mu^2$. Their sum becomes approximately $\mu$-independent over a broad range. At the same time, the factor $D_1(Y_{N_1}-Y_{N_1}^{\rm eq})$ is quasi-static since $D_1\propto K_1$ and $Y_{N_1}-Y_{N_1}^{\rm eq}\sim 1/K_1$, so the freeze-out yield tracks the nearly constant $\varepsilon_{\alpha 1}^{\rm res}=\varepsilon_{{\rm vert},\alpha 1}+\varepsilon_{{\rm self},\alpha 1}^{\rm res}$. This explains the plateau of the freeze-out baryon asymmetry $Y_B^{\rm FO}$ in fig.~\ref{fig:4}.

Quantitatively, we find $\YB^{\rm FO}\gtrsim 10^{-10}$ over the strong-washout regime, $\YB^{\rm FO}\gtrsim 10^{-9}$ across the oscillation-motivated neutrino-mass window $8.65~{\rm meV}\leqslant\tilde{m}_1^{\rm EM}\leqslant 50.1~{\rm meV}$~\cite{Esteban:2024eli}, and a global upper envelope $\YB^{\rm FO}\lesssim 10^{-8}$ (more stringently $\YB^{\rm FO}\lesssim 4\times 10^{-9}$).


\section{Discussion and outlook}
\label{sec:6}

In this section, we extract the physics lessons of our analysis and outline well-motivated avenues for future work. Our treatment is fully EFT-consistent and end-to-end: the UV completion, the one-loop matched coefficient $\CNB$, its RGE running to $\mu_{\rm ref}$, the common dipole scaling of widths and CP sources, and the PU-resummed self-energies evaluated from the same inputs. Flavour resolution is retained throughout to prevent the cancellation of $f_{S_b}$ under flavour summation.

\subsection{Lessons from the EFT pipeline}
\label{sec:6.1}

Two lessons emerge from our EFT-consistent pipeline. First, gauge invariance forces a common dipole scaling on both widths and CP sources, which suppresses non-resonant EMLG. Second, in the quasi-degenerate limit the PU-resummed self-energy counteracts this structural suppression without introducing ad-hoc regulators.

The first of these lessons was established in our previous work (Part~I), which demonstrated a stringent structural suppression in the non-resonant, hierarchical regime: gauge invariance forces a Higgs insertion in $\ONB$, so that after EWSB the broken-phase dipole scales as $\mu_{\alpha i}\propto v/M_{\Psi}^2$~(see eq.~(\ref{eq:2.16})) with an extra one-loop factor $1/(16\pi^2)$ from matching and an $\cO(10\%)$ running effect down to $\mu_{\rm ref}=150~{\rm GeV}$. Consequently, both the flavoured CP sources $\varepsilon_{\alpha i}$ and the washout scale as $\mu^2\propto v^2/M_{\Psi}^4$, which keeps the final baryon asymmetry small in the absence of kinematic enhancement. This mechanism---\enquote{production and washout in lockstep}---was quantified explicitly and provides a useful baseline against which all enhancements should be judged.

Turning to the second lesson, approaching the quasi-degenerate limit unlocks the well-known self-energy enhancement~\cite{FLANZ1995248, FLANZ1996693, COVI1996169, PhysRevD.56.5431} of the loop functions that enter $\varepsilon_{\alpha i}$. In the fully flavoured treatment adopted here, both $f_{S_a}$ and $f_{S_b}$ survive (the latter cancels only after flavour summation). The resonant source can dominate once the mass splitting $\abs{M_1-M_2}$ becomes comparable to the width regulator. The two-body structure $N\to\nu\gamma$, $N\to\nu Z$ remains advantageous for kinetics and keeps the calculation within the same EFT pipeline. Compared with the non-resonant baseline, the qualitative message is that resonant leptogenesis remains a well-established route to achieving a large baryon asymmetry at low scale~\cite{PILAFTSIS2004303}.

In interpreting our numerical results it is important to recall the minimal nature of the EFT adopted here. At the UV level, the Froggatt--Nielsen construction of sec.~\ref{sec:2.2} generates both the SM Dirac Yukawa couplings and the dipole operator $\ONB$, as well as additional $\Delta L=0$ operators. In this second part we have deliberately switched off the Yukawa-induced CP asymmetries and neglected the extra $\Delta L=0$ scatterings in order to isolate the baryon asymmetry that can be attributed to the electromagnetic dipole alone. Because these additional interactions can only increase the washout and do not provide an independent resonant source, the values of $\YB^{\rm FO}$ obtained here should be regarded as optimistic upper bounds within the chosen dipole-dominated EFT.

\subsection{Why the resonant regime yields the observed BAU}
\label{sec:6.2}

In the quasi-degenerate limit, the self-energy denominators produce the width-regularised enhancement of the CP source, while the washout remains controlled by $K_1\propto\sum_{\alpha}\abs{\mu_{\alpha 1}}^2$ (equivalently by $\tilde{m}_1^{\rm EM}$). The net effect is a partial decoupling: the CP source receives an $\cO(1/\Delta M)$ boost until saturated by $\varGamma$, whereas the washout retains its gradual $\tilde{m}_1^{\rm EM}$-dependence. This mechanism mirrors the behaviour known from resonant leptogenesis~\cite{PILAFTSIS2004303} and already noted in the EMLG context~\cite{Choudhury:2011gbi}.

Since we treat $\varepsilon_{e1}^{\rm res}$, $\varepsilon_{\mu 1}^{\rm res}$, and $\varepsilon_{\tau 1}^{\rm res}$ separately, both the $\tilde{f}_{S_a}$ and $\tilde{f}_{S_b}$ self-energy pieces contribute to the CP asymmetries (the flavour sum that would remove $\tilde{f}_{S_b}$ is never taken). This prevents the vanishment that would weaken resonant enhancement.

Within the neutrino-mass window $\tilde{m}_1^{\rm EM}\in [8.65,50.1]~{\rm meV}$~\cite{Esteban:2024eli}, we obtain, in our EFT-consistent framework,
\beq
\label{eq:6.1}
\YB^{\rm FO}\gtrsim 10^{-9},
\eeq
and the maximum over the scan reaches
\beq
\label{eq:6.2}
\YB^{\rm FO}\lesssim 4\times 10^{-9}.
\eeq
Furthermore, over the strong-washout regime, $\YB^{\rm FO}$ is comfortably above $\YB^{\rm obs}\simeq 8.7\times 10^{-11}$~\cite{2020A&A...641A...6P}. The flatness behaviour of $\YB^{\rm FO}$ in the strong-washout regime originates from the saturation of the resonant self-energy source once $\abs{\Delta M}\lesssim\varGamma_2/2$. This behaviour results from the interplay between the Breit--Wigner structure $\mathcal{BW}(\Delta M)$ and the gradual $\tilde{m}_1^{\rm EM}$-dependence of the washout. The detailed mechanism is discussed in sec.~\ref{sec:5.3} and illustrated by the plateau in fig.~\ref{fig:4}.

Part~I showed $\YB^{\rm FO}$ to be suppressed by the common $\mu^2$ scaling of the source and the washout. Here the resonant, flavour-resolved self-energies lift precisely the CP source while leaving the washout scaling intact, thereby opening a wide and stable band of viable leptogenesis. The EFT consistency is preserved throughout: the same one-loop matched $\CNB$ and its RG-evolved value at $\mu_{\rm ref}$ govern both widths and CP asymmetries.

We stress again that the plateau values $\YB^{\rm FO}\gtrsim 10^{-9}$ quoted in this section are obtained within a minimal dipole-dominated setup in which inverse decays are the only washout process. Additional $\Delta L=0$ scatterings induced by the UV completion would only increase the washout and are therefore expected to reduce the final asymmetry $\YB^{\rm FO}$ rather than to enhance it. For this reason we regard the numerical values of $\YB^{\rm FO}$ in the resonant regime as optimistic upper bounds within the chosen EFT framework.

\subsection{Viability at high scales}
\label{sec:6.3}

Ref.~\cite{PhysRevD.78.085024} proposed a gauge-invariant realisation of EMLG in which the dipole operators are already present above EWSB. Heavy Majorana neutrinos $N_k$ couple to the SM via the interaction Lagrangian
\[
\mathcal{L}_{\rm EM}^{(6)}
=-\dfrac{1}{\Lambda^2}\,\mathrm{e}^{-\mathrm{i}\varphi_k/2}\,
\bar{\ell}_j\bigl(\lambda'_{jk}\phi\,\sigma^{\alpha\beta}B_{\alpha\beta}
+\tilde{\lambda}'_{jk}\tau_i\phi\,\sigma^{\alpha\beta}\,W_{\alpha\beta}^i\bigr)
P_RN_k+\text{h.c.},
\]
which respects $\mathrm{SU}(2)_L\times\mathrm{U}(1)_Y$ gauge symmetry. In ref.~\cite{PhysRevD.78.085024}, the authors analysed this system in the symmetric phase, where the relevant processes are three-body decays $N_k\to\ell_j\phi V$ with $V=B,W$ and the associated scatterings. CP violation arises from the interference between the tree-level amplitude and the absorptive part of two-loop vertex and self-energy graphs.

From the EFT point of view, both this high-scale setup and our low-scale resonant scenario are described at $T\ll\Lambda$ by the same dimension-six dipole operator $\ONB$ with a Wilson coefficient $\CNB(\Lambda)$. After running down to the decay scale, all decay widths and CP asymmetries depend on $\CNB$ only through the broken-phase dipole couplings $\mu_{\alpha i}\propto C_{NB,\alpha i}$ defined in eq.~(\ref{eq:2.16}). Schematically, the flavoured CP asymmetries behave as
\beq
\label{eq:6.3}
\varepsilon_{\alpha i}\sim
\dfrac{
\Im\bqty*{\mu_{\alpha i}^{\ast}\mu_{\alpha m}(\mu^{\dagger}\mu)_{im}}}
{\sum_{\beta}\abs{\mu_{\beta i}}^2}
\propto\mu^2
\propto C_{NB}^2,
\eeq
while the total widths scale as $\varGamma_i\propto\sum_{\alpha}\abs{\mu_{\alpha i}}^2\propto C_{NB}^2$, cf.~eqs.~(\ref{eq:2.17})--(\ref{eq:2.22}). Thus, once both descriptions are phrased in terms of the same EFT, the parametric suppression in $\CNB$ is identical for low- and high-scale EMLG: the additional loop in the high-scale analysis of ref.~\cite{PhysRevD.78.085024} corresponds to an extra loop factor in the matching of $\CNB$, but it does not change the overall $C_{NB}^2$ scaling of $\varepsilon_{\alpha i}$.

In our low-scale study we have implemented the resonant enhancement from self-energy graphs in the broken phase, where the dominant processes are the two-body decays $N_i\to\nu_{\alpha}\gamma$, $N_i\to\nu_{\alpha}Z$. In the high-scale realisation of ref.~\cite{PhysRevD.78.085024} the relevant channels are instead three-body decays and scatterings in the symmetric phase, with CP violation arising at effectively two loops in the EFT counting. Whether a quasi-degenerate spectrum at $M_m\gtrsim\cO({\rm TeV})$ can overcome the dimension-six suppression once all thermal effects and the smooth electroweak crossover are treated consistently is a quantitative question that goes beyond the scope of this work. A dedicated high-scale analysis, formulated within the same EFT pipeline and including the temperature-dependent Higgs VEV and flavour decoherence above the crossover, is therefore left to future study.

\subsection{Prospects for other observables}
\label{sec:6.4}

Now that the observed baryon asymmetry can be reproduced, we aim to strengthen the theoretical robustness of low-scale resonant EMLG, mediated by the Wilson coefficient, by further demonstrating its consistency with observations. To this end, let us consider the following observables.

\subsubsection{Light-neutrino masses}
\label{sec:6.4.1}

Radiative corrections involving the dipole operator generically induce light-neutrino-mass terms~\cite{PhysRevD.78.085024, PhysRevLett.95.151802, DAVIDSON2005151, BELL2006377}. The dipole operator induces either Dirac or Majorana masses for the light neutrinos at one loop.
In the Majorana case, the radiative contribution directly provides a light-neutrino mass whose size determines whether an additional seesaw mechanism is required~\cite{Minkowski:1977sc, Yanagida:1979as, Gell-Mann:1979vob, PhysRevLett.44.912}. The analytical expressions for the neutrino mass contributions can be determined precisely within the EFT framework, and the result depends on the UV completion.

\subsubsection{Low-energy observables: $\mu\to e\gamma$, $d_e$, and $\Delta a_{\mu}$}
\label{sec:6.4.2}

At the electroweak scale, our EFT contains the gauge-invariant dipole $\ONB$ with a one-loop matched Wilson coefficient $\CNB$ (see sec.~\ref{sec:2.3}). After EWSB, one may either (i) work directly with the broken-phase dipole couplings $\mu_{\alpha i}$, or (ii) match at the electroweak scale onto the LEFT dipole $\OEG=(\bar{L}\sigma^{\mu\nu}E)HF_{\mu\nu}$ and run within LEFT down to the experimental scales. Here, we adopt (ii) only to fix conventions; the full LEFT running to $m_{\mu}$ and $m_e$ is deferred to Part~III~\cite{Grzadkowski:2010es, Jenkins:2013zja, Jenkins:2013wua, Aebischer:2021uvt}.

The radiative modes $\ell_{\beta}\to\ell_{\alpha}\gamma$, the electron electric dipole moment (eEDM) $d_e$, and $(g-2)_{\ell}$ are dominantly controlled at low energies by $\OEG$. $\OEG$ is induced in two ways that are relevant for EMLG: (a) at one loop by operator mixing from $\ONB$, with the chirality flip provided by a charged-lepton Yukawa insertion on the external leg; and (b) at two loops from two insertions of $\ONB$ (a \enquote{pure-dipole} mechanism). Since $\CNB$ itself is one-loop generated in our UV matching, option (b) is expected to be subleading with respect to (a), owing to the additional loop factor $\sim 1/(16\pi^2)^2$. Nevertheless, both mechanisms can in principle contribute to
the leading effects in the present framework, relevant for $\mathrm{BR}(\mu\to e\gamma)$, $d_e$, and $\Delta a_{\mu}$, and here we only outline the qualitative connection; quantitative correlations will be explored in Part~III.

At low energy,
\beq
\label{eq:6.4}
\begin{split}
&\mathrm{BR}(\mu\to e\gamma)\propto
\abs{(C_{e\gamma})_{e\mu}(m_{\mu})}^2
+\abs{(C_{e\gamma})_{\mu e}(m_{\mu})}^2,\\
&d_e\propto\Im(C_{e\gamma})_{ee}(m_e),\qquad
\Delta a_{\mu}\propto\Re(C_{e\gamma})_{\mu\mu}(m_{\mu}).
\end{split}
\eeq
In Part~III, we will match to LEFT either by running the Wilson coefficients from $\mu_{\rm ref}=150~{\rm GeV}$ down to $\mu=m_{\mu},m_e$ according to the RGE of QED, or by starting directly at $\mu=m_Z\simeq 91.2~{\rm GeV}$. We then run down to $\mu_{\rm low}\simeq m_{\mu},m_e$ for the respective observables~\cite{Crivellin:2013hpa, Aebischer:2021uvt}.

The resonant enhancement that lifts the BAU arises from right-handed-neutrino self-energy contributions in CP-odd decay asymmetries (width-regularised poles). By contrast, $\CEG(q^2=0)$ corresponds to an analytic on-shell photon vertex: there is no width-regularised pole in $\CEG$. Therefore, BAU and low-energy dipoles do not share the resonance.

\subsection{Open theoretical issues}
\label{sec:6.5}

Several ingredients deserve dedicated follow-up:

(i) Thermal widths and masses, and the electroweak crossover. A thermal treatment of the regulator and of the on-shell conditions in the Cutkosky cuts~\cite{10.1063/1.1703676}, together with a realistic implementation of the temperature-dependent Higgs VEV $v(T)$ near the electroweak symmetry breaking, would reduce theoretical systematics around the resonance. In particular, incorporating the lattice-based determinations of the sphaleron rate and of the crossover profile~\cite{PhysRevLett.113.141602, PhysRevD.93.025003} within our EFT pipeline would allow for a more accurate assessment of the temperature range over which the BAU is generated.

(ii) Scatterings. We included inverse decays as the dominant washout; adding the $2\leftrightarrow 2$ scatterings induced by the dipole operator $\ONB$---both lepton-number-violating and lepton-number-conserving channels such as $N_i\gamma\leftrightarrow\nu_{\alpha}\gamma$---would quantify their impact on the strong-washout regime of the peak. Since gauge bosons are introduced, we need to develop a new implementation of real-intermediate-state (RIS) subtraction~\cite{PILAFTSIS2004303} within the Boltzmann equation or the flavour covariant transport equation, whereas in the closed-time-path (CTP/Schwinger--Keldysh) formalism with Kadanoff--Baym equations~\cite{BENEKE20101,PhysRevD.80.125027,PhysRevD.108.096034} the RIS subtraction becomes unnecessary. At the present low scale, scatterings in which gauge bosons participate are expected to correct the strong-washout plateau.

(iii) Symmetric-phase dynamics and higher-multiplicity processes. Our present analysis focuses on the broken phase and on the two-body decays and inverse decays induced by $\ONB$. A complete treatment should also include the symmetric-phase decays and scatterings generated by the same operator, notably $1\to 3$ processes and related $2\leftrightarrow 2$ channels, consistently across the electroweak crossover. Extending the EFT pipeline to cover these additional processes, and matching the symmetric- and broken-phase descriptions in a unified way, is an important step toward a fully quantitative theory of low-scale EMLG.

(iv) RG improvement. Our one-loop evolution of $\CNB$ already captures the numerically important effects; nevertheless, a two-loop study and matching-scale variation would provide a robust error budget.

(v) Initial conditions. While we assumed a thermal $N_1$ abundance, the resonant window is sensitive to departures from equilibrium. Exploring partially populated (\enquote{freeze-in-like}) initial conditions, where $N_1$ is underabundant but nonzero at $T\simeq T_{\rm EW}$, would be worthwhile.

(vi) Momentum- and helicity-resolved quantum kinetic equations. In this work we have evolved the right-handed neutrino ensemble using momentum-averaged densities with Maxwell--Boltzmann statistics and a single temperature scale, which we expect to be adequate in the regime $M_1/T\gtrsim 3$ relevant for our benchmark. A more refined treatment would couple the right-handed neutrinos and lepton doublets through quantum kinetic equations (QKE) in momentum space, including helicity dependence and medium induced modifications of the dispersion relations, along the lines of recent freeze-in leptogenesis studies~\cite{Ghiglieri:2017gjz}. Implementing such equations consistently within our EFT pipeline would significantly increase the technical complexity, and we leave this as a key direction for future work.



\section{Conclusions}
\label{sec:7}

In summary, our EFT-consistent analysis of low-scale, resonant electromagnetic leptogenesis (EMLG) leads to the following points:

(i) Structural suppression and its remedy. Gauge invariance implies that the electromagnetic dipole interaction originates only at dimension-six, through the operator $\ONB=(\bar{L}\sigma^{\mu\nu}N)\tilde{H}B_{\mu\nu}$. After EWSB, the broken-phase dipole couplings obey $\mu_{\alpha i}\propto v/M_{\Psi}^2$, so both total widths and CP-odd sources scale as $v^2/M_{\Psi}^4$ (with an additional one-loop suppression from matching and running). This universal scaling---independent of flavour and spectrum---accounts for the failure of non-resonant EMLG within the EFT framework. In the quasi-degenerate limit, however, self-energy-driven enhancement overcomes this structural suppression by resurrecting the CP source.

(ii) Successful resonant regime. When $M_1\simeq M_2$, the flavour-resolved self-energy loop functions become resonant; because we never perform the flavour sum, the antisymmetric $f_{S_b}$ piece does not cancel and can dominate. Across the parameter ranges scanned, the freeze-out baryon asymmetry satisfies~\cite{2020A&A...641A...6P}
\beq
\label{eq:7.1}
\begin{split}
\YB^{\rm FO}&\gtrsim 10^{-10}\gtrsim\YB^{\rm obs},
\quad(\text{strong-washout regime}),\\
\YB^{\rm FO}&\gtrsim 10^{-9},
\hspace{4.7em}(\text{neutrino-mass window}),
\end{split}
\eeq
and reaches a maximum $\YB^{\rm FO}\lesssim 10^{-8}$. This flat, high-yield behaviour is traced to the saturation of the width-regularised self-energy source together with the gradual $K_1$-driven washout.

Finally, a systematic confrontation with other observables is now well motived. In particular, (a) light-neutrino masses radiatively induced by the dipole operator, and (b) low-energy probes $\mu\to e\gamma$, $d_e$, and $\Delta a_{\mu}$, analysed within a consistent matching to LEFT and RG flow down to experimental scales, will sharpen the viability of low-scale resonant EMLG and delineate its testable parameter space. This programme naturally bridges the present work to the next stage of electromagnetic leptogenesis.

\acknowledgments

The author is grateful to K.~Hotokezaka, R.~Jinno, and R.~Namba for helpful comments on the manuscript.

\appendix
\section{Computation of self-energy contributions}
\label{sec:A}

\enlargethispage{\baselineskip}

For clarity, in this appendix we restrict the computation of the self-energy contributions to the photon channel. In order to obtain the Breit--Wigner--form loop functions in eq.~(\ref{eq:3.5}), we Dyson-resum the internal propagator of $N_m$ as in eq.~(\ref{eq:3.3}), and subsequently compute the Feynman graphs, which are shown in fig.~\ref{fig:5}. The corresponding non-resonant calculation can be found in~\cite{Law:2008yyq}.

\subsection{The contribution of graph~(a)}
\label{sec:A.1}

\begin{figure}[t]
\centering
\begin{minipage}[b]{0.49\hsize}
\centering
\begin{tikzpicture}[baseline=-0.1cm,scale=1]
\begin{feynhand}
\tikzset{
    wavy/.style={decorate, decoration={snake, amplitude=0.75mm, segment length=2.9mm}, thick},
}
\setlength{\feynhandblobsize}{3mm}
\vertex (i1) at (0,0);
\coordinate (w2) at (1.5,0);
\coordinate (w3) at (3.5,0);
\coordinate (w4) at (5.5,0);
\vertex (f5) at (7,1.5);
\vertex (f6) at (7,-1.5);
\propag [plain] (i1) to [edge label=$N_i(k)$] (w2);
\propag [antfer] (w2) to [half right, looseness=1.75, edge label'=$\nu_{\beta}(q_1)$] (w3);
\draw[wavy]
      (w2) arc[start angle=180,end angle=0,radius=1cm]
      node[midway,above] {$\gamma_{\rho}(q_2)$};
\propag [plain] (w3) to [edge label=$N_m(k)$] (w4);
\propag [fer] (w4) to [edge label=$\nu_{\alpha}(p)$] (f5);
\draw [wavy] (w4) to [edge label'=$\gamma_{\sigma}(q)$] (f6);
\node at (1,-0.4) {$\tilde{V}_{\beta i}^{\rho}$};
\node at (4.15,-0.4) {$V_{\beta m}^{\mu}$};
\node at (6.25,0) {$V_{\alpha m}^{\sigma}$};
\node at (0,2) {\bfseries (a)};
\vertex [grayblob] (w2) at (1.5,0) {};
\vertex [grayblob] (w3) at (3.5,0) {};
\vertex [grayblob] (w4) at (5.5,0) {};
\end{feynhand}
\end{tikzpicture}
\end{minipage}
\begin{minipage}[b]{0.49\hsize}
\centering
\begin{tikzpicture}[baseline=-0.1cm,scale=1]
\begin{feynhand}
\tikzset{
    wavy/.style={decorate, decoration={snake, amplitude=0.75mm, segment length=2.9mm}, thick},
}
\setlength{\feynhandblobsize}{3mm}
\vertex (i1) at (0,0);
\coordinate (w2) at (1.5,0);
\coordinate (w3) at (3.5,0);
\coordinate (w4) at (5.5,0);
\vertex (f5) at (7,1.5);
\vertex (f6) at (7,-1.5);
\propag [plain, mom'={$k$}] (i1) to [edge label=$i$] (w2);
\propag [antfer, mom'={$q_1$}] (w2) to [half right, looseness=1.75] (w3);
\draw[wavy] 
	(w2) arc[start angle=180,end angle=0,radius=1cm];
\draw[-{Stealth[length=3.25pt,width=2.25pt]},
      line width=0.3pt]
      (2.5,0) ++(135:1.3cm)
      arc[start angle=135, delta angle=-90, radius=1.3cm];
\node at (2.5,1.6) {$q_2$};
\propag [plain, mom'={$k$}] (w3) to (w4);
\propag [fer, mom={$p$}] (w4) to [edge label'=$\alpha$] (f5);
\draw [wavy] (w4) to [edge label=$\sigma$] (f6);
\draw [draw=none, mom'={[arrow distance=6pt] {$q$}}] (w4) to (f6);
\node at (2,-0.55) {$\beta$};
\node at (3,-0.55) {$\beta$};
\node at (2,0.6) {$\rho$};
\node at (3,0.6) {$\mu$};
\node at (4,0.3) {$m$};
\node at (5,0.3) {$m$};
\vertex [grayblob] (w2) at (1.5,0) {};
\vertex [grayblob] (w3) at (3.5,0) {};
\vertex [grayblob] (w4) at (5.5,0) {};
\end{feynhand}
\end{tikzpicture}
\end{minipage}
\begin{minipage}[b]{0.49\hsize}
\centering
\begin{tikzpicture}[baseline=-0.1cm,scale=1]
\begin{feynhand}
\tikzset{
    wavy/.style={decorate, decoration={snake, amplitude=0.75mm, segment length=2.9mm}, thick},
}
\setlength{\feynhandblobsize}{3mm}
\vertex (i1) at (0,0);
\coordinate (w2) at (1.5,0);
\coordinate (w3) at (3.5,0);
\coordinate (w4) at (5.5,0);
\vertex (f5) at (7,1.5);
\vertex (f6) at (7,-1.5);
\propag [plain] (i1) to [edge label=$N_i(k)$] (w2);
\propag [fer] (w2) to [half right, looseness=1.75, edge label'=$\nu_{\beta}(q_1)$] (w3);
\draw[wavy]
      (w2) arc[start angle=180,end angle=0,radius=1cm]
      node[midway,above] {$\gamma_{\rho}(q_2)$};
\propag [plain] (w3) to [edge label=$N_m(k)$] (w4);
\propag [fer] (w4) to [edge label=$\nu_{\alpha}(p)$] (f5);
\draw [wavy] (w4) to [edge label'=$\gamma_{\sigma}(q)$] (f6);
\node at (1,-0.4) {$V_{\beta i}^{\rho}$};
\node at (4.15,-0.4) {$\tilde{V}_{\beta m}^{\mu}$};
\node at (6.25,0) {$V_{\alpha m}^{\sigma}$};
\node at (0,2) {\bfseries (b)};
\vertex [grayblob] (w2) at (1.5,0) {};
\vertex [grayblob] (w3) at (3.5,0) {};
\vertex [grayblob] (w4) at (5.5,0) {};
\end{feynhand}
\end{tikzpicture}
\end{minipage}
\begin{minipage}[b]{0.49\hsize}
\centering
\begin{tikzpicture}[baseline=-0.1cm,scale=1]
\begin{feynhand}
\tikzset{
    wavy/.style={decorate, decoration={snake, amplitude=0.75mm, segment length=2.9mm}, thick},
}
\setlength{\feynhandblobsize}{3mm}
\vertex (i1) at (0,0);
\coordinate (w2) at (1.5,0);
\coordinate (w3) at (3.5,0);
\coordinate (w4) at (5.5,0);
\vertex (f5) at (7,1.5);
\vertex (f6) at (7,-1.5);
\propag [plain, mom'={$k$}] (i1) to [edge label=$i$] (w2);
\propag [fer, mom'={$q_1$}] (w2) to [half right, looseness=1.75] (w3);
\draw[wavy] 
	(w2) arc[start angle=180,end angle=0,radius=1cm];
\draw[-{Stealth[length=3.25pt,width=2.25pt]},
      line width=0.3pt]
      (2.5,0) ++(135:1.3cm)
      arc[start angle=135, delta angle=-90, radius=1.3cm];
\node at (2.5,1.6) {$q_2$};
\propag [plain, mom'={$k$}] (w3) to (w4);
\propag [fer, mom={$p$}] (w4) to [edge label'=$\alpha$] (f5);
\draw [wavy] (w4) to [edge label=$\sigma$] (f6);
\draw [draw=none, mom'={[arrow distance=6pt] {$q$}}] (w4) to (f6);
\node at (2,-0.55) {$\beta$};
\node at (3,-0.55) {$\beta$};
\node at (2,0.6) {$\rho$};
\node at (3,0.6) {$\mu$};
\node at (4,0.3) {$m$};
\node at (5,0.3) {$m$};
\vertex [grayblob] (w2) at (1.5,0) {};
\vertex [grayblob] (w3) at (3.5,0) {};
\vertex [grayblob] (w4) at (5.5,0) {};
\end{feynhand}
\end{tikzpicture}
\end{minipage}
\caption{The two types of Feynman graphs (a) and (b) for the self-energy contributions in EMLG. In each case, the left panel shows the particle species involved in the reaction, the four-momenta, and the vertices, while the right panel indicates the directions of the four-vectors $k,p,q,q_1,q_2$.}
\label{fig:5}
\end{figure}
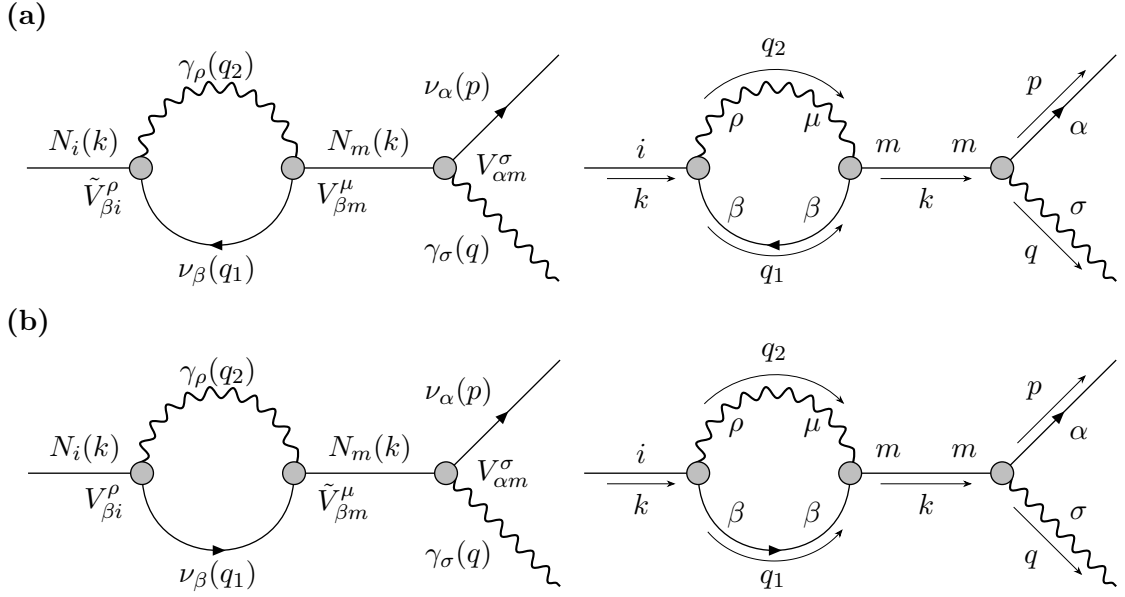

Let us first compute graph (a) in fig.~\ref{fig:5}. The propagators of $\nu$, $N_m$, and $\gamma$ are given, with $S_{N_m}$ being Dyson-resummed right-handed-neutrino propagator as in eq.~(\ref{eq:3.3}):
\begin{equation}
\label{eq:A.1}
S_{\nu}(q_1)\simeq
\dfrac{\mathrm{i}\slashed{q}_1}{q_1^2+\mathrm{i}\epsilon},\qquad
S_{N_m}(k)
=\dfrac{\mathrm{i}(\slashed{k}+M_m)}
{k^2-M_m^2+\mathrm{i}M_m\varGamma_m+\mathrm{i}\epsilon},
\qquad
D_{\mu\nu}(q_2)
=\dfrac{-\mathrm{i}g_{\mu\nu}}{q_2^2+\mathrm{i}\epsilon}.
\end{equation}
Moreover, the three effective dipole vertices in fig.~\ref{fig:5}~(a) read
\begin{align}
\tilde{V}_{\beta i}^{\rho}(q_2)
&=2\mu_{\beta i}^{\ast}\sigma^{\rho\lambda}q_{2\lambda}P_L,
\label{eq:A.2}\\
V_{\beta m}^{\mu}(-q_2)
&=2\mu_{\beta m}\sigma^{\mu\eta}(-q_{2\eta})P_R,
\label{eq:A.3}\\
V_{\alpha m}^{\sigma}(q)
&=2\mu_{\alpha m}\sigma^{\sigma\kappa}q_{\kappa}P_R.
\label{eq:A.4}
\end{align}
Accordingly, we define
\beq
\label{eq:A.5}
A_{\mu}^{(5)}\coloneqq
\mu_{\alpha i}^{\ast}\mu_{\alpha m}\mu_{\beta m}\mu_{\beta i}^{\ast}.
\eeq
The interference term corresponding to graph (a) is obtained by
\begin{align}
I_{\rm self}^{\rm (a)}
&=\dfrac{1}{2}\sum_{s,s'}\sum_{\rm pol.}
\bqty*{\mathrm{i}\mathcal{M}_{\gamma}^{\rm tree}(s,s')}^{\dagger}
\bqty*{\mathrm{i}\mathcal{M}_{\gamma}^{\text{\scriptsize 1-loop}}(s,s')}
\notag\\
&=\dfrac{1}{2}\sum_{s,s'}\sum_{\rm pol.}
\bqty*{2\mu_{\alpha i}^{\ast}\varepsilon_{\nu}(q)\bar{u}_i(k)
\sigma^{\nu\chi}q_{\chi}P_Lu_{\alpha}(p)}\notag\\
&\quad\times
\int\!\!\dfrac{\dd^4q_1}{(2\pi)^4}\;
\bar{u}_{\alpha}(p)V_{\alpha m}^{\sigma}(q)\varepsilon_{\sigma}^{\ast}(q)
S_{N_m}(k)V_{\beta m}^{\mu}(-q_2)
S_{\nu_{\beta}}(-q_1)\tilde{V}_{\beta i}^{\rho}(q_2)
D_{\mu\rho}(q_2)u_i(k)\notag\\
&=\mathrm{i}A_{\mu}^{(5)}M_iM_m
\int\!\!\dfrac{\dd^4q_1}{(2\pi)^4}\;
\dfrac{(p\cdot q)[-256(q\cdot q_2)(q_1\cdot q_2)+64(q\cdot q_1)q_2^2]}
{(k^2-M_m^2+\mathrm{i}M_m\varGamma_m+\mathrm{i}\epsilon)
(q_1^2+\mathrm{i}\epsilon)(q_2^2+\mathrm{i}\epsilon)}.
\label{eq:A.6}
\end{align}
Considering Denner's fermion-flow prescription~\cite{DENNER1992467} (see Part~I for details), the spinor chain
\beq
\label{eq:A.7}
\bar{u}_{\alpha}V_{\alpha m}^{\sigma}(q)S_{N_m}(k)
V_{\beta m}^{\mu}(-q_2)S_{\nu_{\beta}}(-q_1)
\tilde{V}_{\beta i}^{\rho}(q_2)u_i
\eeq
has all three vertices read along the flow, so that no reversed vertex appears and all Dirac structures are of the form. Consequently, no additional minus sign arises.

\begin{figure}[t]
\centering
\begin{tikzpicture}[baseline=-0.1cm,scale=1.2]
\begin{feynhand}
\tikzset{
    wavy/.style={decorate, decoration={snake, amplitude=0.75mm, segment length=2.9mm}, thick},
}
\setlength{\feynhandblobsize}{3mm}
\vertex (i1) at (0,0);
\coordinate (w2) at (1.5,0);
\coordinate (w3) at (3.5,0);
\coordinate (w4) at (5.5,0);
\vertex (f5) at (7,1.25);
\vertex (f6) at (7,-1.25);
\vertex (w7) at (2.8,1.5);
\vertex (w8) at (2.8,-1.5);
\propag [plain] (i1) to (w2);
\propag [antfer] (w2) to [half right, looseness=1.75] (w3);
\draw[wavy]
      (w2) arc[start angle=180,end angle=0,radius=1cm];
\propag [plain] (w3) to (w4);
\propag [fer] (w4) to (f5);
\draw [wavy] (w4) to (f6);
\propag [ghost] (w7) to (w8);
\vertex [grayblob] (w2) at (1.5,0) {};
\vertex [grayblob] (w3) at (3.5,0) {};
\vertex [grayblob] (w4) at (5.5,0) {};
\end{feynhand}
\end{tikzpicture}
\caption{Cutkosky cutting rules~\cite{10.1063/1.1703676} for the self-energy contribution. The vertical dotted line indicates the cut, which traverses both the photon and neutrino lines forming the loop. Since no other propagators enter the loop, this combination is the only valid Cutkosky cut.}
\label{fig:6}
\end{figure}
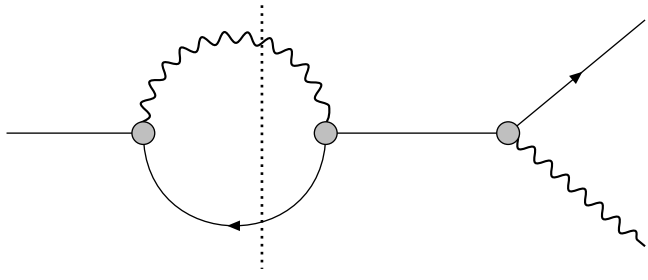

To evaluate the discontinuity of eq.~(\ref{eq:A.6}), we apply the Cutkosky rules~\cite{10.1063/1.1703676}. In fig.~\ref{fig:6}, the only possible cut that simultaneously puts the one-loop momenta on shell is the vertical cut through the fermion line $q_1$ and the photon line $q_2$. According to the Cutkosky rules~\cite{10.1063/1.1703676}
\begin{equation}
\label{eq:A.8}
\dfrac{1}{p^2-m^2+\mathrm{i}\epsilon}
\to -2\pi\mathrm{i}\,\delta(p^2-m^2)\,\Theta(p^0),
\end{equation}
the denominators in eq.~(\ref{eq:A.6}) transform as
\begin{align}
\dfrac{1}{q_1^2+\mathrm{i}\epsilon}
&\to -2\pi\mathrm{i}\,\delta(q_1^2)\Theta(E_1),
\label{eq:A.9}\\
\dfrac{1}{q_2^2+\mathrm{i}\epsilon}
&\to -2\pi\mathrm{i}\,\delta(q_2^2)\Theta(E_2)
=-2\pi\mathrm{i}\,\delta\bigl((k-q_1)^2\bigr)\Theta(M_i-E_1).
\label{eq:A.10}
\end{align}
Here, $q_1=(E_1,\bm{q}_1)$, $q_2=(E_2,\bm{q}_2)=k-q_1=(M_i-E_1,-\bm{q}_1)$. Furthermore, setting
\begin{equation}
\label{eq:A.11}
\begin{split}
&k=(M_i,\bm{0}),\quad
p=\pqty*{\dfrac{M_i}{2},-\bm{q}},\quad
q=\pqty*{\dfrac{M_i}{2},\bm{q}},\quad
\norm{\bm{q}}=\dfrac{M_i}{2},\\[0.25em]
&k\cdot p=p\cdot q=q\cdot k=\dfrac{M_i^2}{2},\quad
k^2=M_i^2,\quad
p^2\simeq 0,\quad
q^2=0
\end{split}
\end{equation}
and taking the limit $\epsilon\to 0$, the discontinuity of $J_{\rm self}^{\rm (a)}\coloneqq I_{\rm self}^{\rm (a)}/A_{\mu}^{(5)}$ is obtained as
\begin{align}
\Disc(J_{\rm self}^{\rm (a)})
&=\mathrm{i}M_iM_m
\int\!\!\dfrac{\dd^4q_1}{(2\pi)^4}\;
\dfrac{(-2\pi\mathrm{i})^2\delta(q_1^2)\delta\bigl((k-q_1)^2\bigr)
\Theta(E_1)\Theta(M_i-E_1)}{k^2-M_m^2+\mathrm{i}M_m\varGamma_m}
\notag\\
&\quad\times
M_i^2\bqty*{-128(q\cdot q_2)(q_1\cdot q_2)+32(q\cdot q_1)q_2^2}\notag\\
&=-\dfrac{\mathrm{i}M_i^4M_m}
{4\pi^2(M_i^2-M_m^2+\mathrm{i}M_m\varGamma_m)}\notag\\
&\quad\times
\int\dd^3q_1\;\dd E_1\;\dfrac{1}{2\norm{\bm{q}_1}}
\delta(E_1-\norm{\bm{q}_1})
\delta\bigl((M_i-E_1)^2-\norm{\bm{q}_1}^2\bigr)
\Theta(E_1)\Theta(M_i-E_1)\notag\\
&\quad\times
\bigl[-64(M_i-E_1+\norm{\bm{q}_1}\cos\theta)
(M_iE_1-E_1^2+\norm{\bm{q}_1}^2)\notag\\
&\hspace{8em}
+16(E_1-\norm{\bm{q}_1}\cos\theta)
\bigl((M_i-E_1)^2-\norm{\bm{q}_1}^2\bigr)\bigr],
\label{eq:A.12}
\end{align}
where we used
\begin{align}
q\cdot q_2
&=q\cdot (k-q_1)
=\dfrac{M_i^2}{2}-\dfrac{M_i}{2}(E_1-\norm{\bm{q}_1}\cos\theta),
\label{eq:A.13}\\[0.25em]
q_1\cdot q_2
&=q_1\cdot (k-q_1)
=M_iE_1-E_1^2+\norm{\bm{q}_1}^2.
\label{eq:A.14}
\end{align}

Performing the $E_1$ integral using $\delta\bigl((M_i-E_1)^2-\norm{\bm{q}_1}^2\bigr)$ fixes $E_1=\norm{\bm{q}_1}$. The remaining angular integrals are carried out in the rest frame of $N_i$ with the angle $\theta$ between $\bm{q}$ and $\bm{q}_1$, yielding
\begin{align}
\Disc(J_{\rm self}^{\rm (a)})
&=-\dfrac{\mathrm{i}M_i^4M_m}
{8\pi^2(M_i^2-M_m^2+\mathrm{i}M_m\varGamma_m)}\notag\\
&\quad\times
\int\norm{\bm{q}_1}^2\;\dd\norm{\bm{q}_1}\;\dd\varOmega\;
\dfrac{1}{\norm{\bm{q}_1}}\dfrac{1}{\abs{-2M_i}}
\delta(M_i-2\norm{\bm{q}_1})
\Theta(M_i-\norm{\bm{q}_1})\notag\\
&\quad\times
\bigl[-64(M_i-\norm{\bm{q}_1}+\norm{\bm{q}_1}\cos\theta)
M_i\norm{\bm{q}_1}
+16\norm{\bm{q}_1}(1-\cos\theta)
M_i(M_i-2\norm{\bm{q}_1})\bigr]\notag\\
&=\dfrac{2\mathrm{i}M_i^7M_m
(M_i^2-M_m^2-\mathrm{i}M_m\varGamma_m)}
{\pi\bqty*{(M_i^2-M_m^2)^2+M_m^2\varGamma_m^2}}.
\label{eq:A.15}
\end{align}
Consequently, the imaginary part of the interference term corresponding to fig.~\ref{fig:5}~(a) is
\begin{align}
\Im(J_{\rm self}^{\rm (a)})
&=\dfrac{1}{2\mathrm{i}}\Disc(J_{\rm self}^{\rm (a)})
=\dfrac{M_i^7M_m}{\pi}\,
\dfrac{M_i^2-M_m^2}{(M_i^2-M_m^2)^2+M_m^2\varGamma_m^2},
\label{eq:A.16}
\end{align}
from which the contribution of graph (a) to the self-energy part of the CP asymmetry reads
\begin{align}
\varepsilon_{{\rm self},\alpha i}^{\gamma,{\rm (a)}}
&=-\dfrac{4}{\varGamma_i}\Im(A_{\mu}^{(5)})
\Im(J_{\rm self}^{\rm (a)}\varPhi_2)\notag\\
&=-\dfrac{M_i^2}{2\pi\sum_{\beta}\abs{\mu_{\beta i}}^2}\,
\Im\bqty*{\mu_{\alpha i}^{\ast}\mu_{\alpha m}(\mu^{\dagger}\mu)_{im}}\,
\dfrac{(M_i^2-M_m^2)M_iM_m}{(M_i^2-M_m^2)^2+M_m^2\varGamma_m^2},
\label{eq:A.17}
\end{align}
where $\varPhi_2=\norm{\bm{q}}/(8\pi M_i^2)$ denotes the two-body phase-space factor for each (anti)neutrino final state~\cite{Kugo1989I, Schwartz_2013}. Rewriting the Breit--Wigner form in terms of $x=M_m^2/M_i^2$, one has $M_iM_m\simeq M_i^2\sqrt{x}$ and $M_i^2-M_m^2=M_i^2(1-x)$, which yields
\begin{align}
\tilde{f}_{S_a}(x)&\coloneqq
\dfrac{(M_i^2-M_m^2)M_iM_m}{(M_i^2-M_m^2)^2+M_m^2\varGamma_m^2}
\notag\\[0.25em]
&=\dfrac{\sqrt{x}\,(1-x)}{(1-x)^2+x(\varGamma_m/M_i)^2}
\simeq\dfrac{\sqrt{x}\,(1-x)}{(1-x)^2+(\varGamma_m/M_i)^2},
\label{eq:A.18}
\end{align}
where the last step holds in the quasi-degenerate limit $x\simeq 1$. This is
the Breit--Wigner envelope (up to the slowly varying factor $\sqrt{x}$\,) for the self-energy piece regularised by the total width $\varGamma_m$.

\subsection{The contribution of graph~(b)}
\label{sec:A.2}

Next, let us compute graph~(b) in fig.~\ref{fig:5}. In this case, the three vertices read
\begin{align}
V_{\beta i}^{\rho}(q_2)
&=2\mu_{\beta i}\sigma^{\rho\lambda}q_{2\lambda}P_R,
\label{eq:A.19}\\
\tilde{V}_{\beta m}^{\mu}(-q_2)
&=2\mu_{\beta m}^{\ast}\sigma^{\mu\eta}(-q_{2\eta})P_L,
\label{eq:A.20}\\
V_{\alpha m}^{\sigma}(q)
&=2\mu_{\alpha m}\sigma^{\sigma\kappa}q_{\kappa}P_R.
\label{eq:A.21}
\end{align}
Accordingly, the interference term corresponding to graph (b), defining
\beq
\label{eq:A.22}
B_{\mu}^{(5)}\coloneqq
\mu_{\alpha i}^{\ast}\mu_{\alpha m}\mu_{\beta m}^{\ast}\mu_{\beta i},
\eeq
and we obtain
\begin{align}
I_{\rm self}^{\rm (b)}
&=\dfrac{1}{2}\sum_{s,s'}\sum_{\rm pol.}
\bqty*{\mathrm{i}\mathcal{M}_{\gamma}^{\rm tree}(s,s')}^{\dagger}
\bqty*{\mathrm{i}\mathcal{M}_{\gamma}^{\text{\scriptsize 1-loop}}(s,s')}
\notag\\
&=\dfrac{1}{2}\sum_{s,s'}\sum_{\rm pol.}
\bqty*{2\mu_{\alpha i}^{\ast}\varepsilon_{\nu}(q)\bar{u}_i(k)
\sigma^{\nu\chi}q_{\chi}P_Lu_{\alpha}(p)}\notag\\
&\quad\times
\int\!\!\dfrac{\dd^4q_1}{(2\pi)^4}\;
\bar{u}_{\alpha}(p)V_{\alpha m}^{\sigma}(q)\varepsilon_{\sigma}^{\ast}(q)
S_{N_m}(k)\tilde{V}_{\beta m}^{\mu}(-q_2)
S_{\nu_{\beta}}(q_1)V_{\beta i}^{\rho}(q_2)
D_{\mu\rho}(q_2)u_i(k)\notag\\
&=\dfrac{1}{2}\sum_{s,s'}
\bqty*{\sum_{\rm pol.}\varepsilon_{\nu}(q)\varepsilon_{\sigma}^{\ast}(q)}
\bqty*{2\mu_{\alpha i}^{\ast}\bar{u}_i(k)
\sigma^{\nu\chi}q_{\chi}P_Lu_{\alpha}(p)}\notag\\
&\quad\times
\int\!\!\dfrac{\dd^4q_1}{(2\pi)^4}\;
\bar{u}_{\alpha}(p)V_{\alpha m}^{\sigma}(q)
S_{N_m}(k)\tilde{V}_{\beta m}^{\mu}(-q_2)
S_{\nu_{\beta}}(q_1)V_{\beta i}^{\rho}(q_2)
D_{\mu\rho}(q_2)u_i(k)
\label{eq:A.23}
\end{align}
According to Denner's fermion-flow prescription~\cite{DENNER1992467}, in the spinor chain
\beq
\label{eq:A.24}
\bar{u}_{\alpha}V_{\alpha m}^{\sigma}(q)S_{N_m}(k)
\tilde{V}_{\beta m}^{\mu}(-q_2)S_{\nu_{\beta}}(q_1)
V_{\beta i}^{\rho}(q_2)u_i,
\eeq
the reversed vertex is $\tilde{V}_{\beta m}(-q_2)$. The Dirac structure of the reversed vertex is
$\Gamma^{\rm rev}=\eta_{\sigma}\Gamma=-\Gamma$, from which an additional sign factor $-1$ arises. Taking this into account, one finds
\begin{align}
I_{\rm self}^{\rm (b)}
&=(-1)\times(-1)\mu_{\alpha i}^{\ast}g_{\nu\sigma}
\bqty*{\sum_{s}u_i(k)\bar{u}_i(k)}
\sigma^{\nu\chi}q_{\chi}P_L\notag\\
&\quad\times
\int\!\!\dfrac{\dd^4q_1}{(2\pi)^4}\;
\bqty*{\sum_{s'}u_{\alpha}(p)\bar{u}_{\alpha}(p)}
(2\mu_{\alpha m}\sigma^{\sigma\kappa}q_{\kappa}P_R)
\dfrac{\mathrm{i}(\slashed{k}+M_m)}
{k^2-M_m^2+\mathrm{i}M_m\varGamma_m+\mathrm{i}\epsilon}
\notag\\
&\quad\times
[2\mu_{\beta m}^{\ast}\sigma^{\mu\eta}(-q_{2\eta})P_L]
\dfrac{\mathrm{i}\slashed{q}_1}{q_1^2+\mathrm{i}\epsilon}
(2\mu_{\beta i}\sigma^{\rho\lambda}q_{2\lambda}P_R)
\dfrac{-\mathrm{i}g_{\mu\rho}}{q_2^2+\mathrm{i}\epsilon}\notag\\
&=32\mathrm{i}M_i^4B_{\mu}^{(5)}
\int\!\!\dfrac{\dd^4q_1}{(2\pi)^4}\notag\\
&\quad\times
\dfrac{4(q\cdot q_2)(q_1\cdot q_2)-(q\cdot q_1)q_2^2
-4(k\cdot q_2)(q_1\cdot q_2)+(k\cdot q_1)q_2^2}
{(k^2-M_m^2+\mathrm{i}M_m\varGamma_m+\mathrm{i}\epsilon)(q_1^2+\mathrm{i}\epsilon)
(q_2^2+\mathrm{i}\epsilon)}.
\label{eq:A.25}
\end{align}


The discontinuity of the integral (\ref{eq:A.25}) follows from the Cutkosky cut shown in fig.~\ref{fig:6}, and setting $q_1=(E_1,\bm{q}_1)$ and $q_2=k-q_1$ yields
\begin{align}
\Disc(J_{\rm self}^{\rm (b)})
&=\dfrac{32\mathrm{i}M_i^4}{M_i^2-M_m^2+\mathrm{i}M_m\varGamma_m}
\int\!\!\dfrac{\dd^4q_1}{(2\pi)^4}\;
(-2\pi\mathrm{i})^2\delta(q_1^2)\delta\bigl((k-q_1)^2\bigr)
\Theta(E_1)\Theta(M_i-E_1)\notag\\
&\hspace{7em}\times
\bqty*{4(q\cdot q_2)(q_1\cdot q_2)-(q\cdot q_1)q_2^2
-4(k\cdot q_2)(q_1\cdot q_2)+(k\cdot q_1)q_2^2}\notag\\
&=-\dfrac{8\mathrm{i}M_i^4}
{\pi^2(M_i^2-M_m^2+\mathrm{i}M_m\varGamma_m)}
\int\norm{\bm{q}_1}^2\;\dd\norm{\bm{q}_1}\;\dd\varOmega\;\dd E_1\notag\\
&\qquad\times
\delta(E_1^2-\norm{\bm{q}_1}^2)
\delta\bigl((M_i-E_1)^2-\norm{\bm{q}_1}^2\bigr)
\Theta(E_1)\Theta(M_i-E_1)\notag\\
&\qquad\times
\biggl\{4(M_iE_1-E_1^2+\norm{\bm{q}_1}^2)
\bqty*{\dfrac{M_i^2}{2}-\dfrac{M_i}{2}(E_1-\norm{\bm{q}_1}\cos\theta)
-M_i^2+M_iE_1}\notag\\
&\hspace{9em}
+\bqty*{M_iE_1-\dfrac{M_i}{2}(E_1-\norm{\bm{q}_1}\cos\theta)}
\bqty*{(M_i-E_1)^2-\norm{\bm{q}_1}^2}\biggr\}.
\label{eq:A.26}
\end{align}
Carrying out first the $E_1$ integral and subsequently the $\norm{\bm{q}_1}$ integral, one finds
\begin{align}
\Disc(J_{\rm self}^{\rm (b)})
&=-\dfrac{4\mathrm{i}M_i^5}
{\pi^2(M_i^2-M_m^2+\mathrm{i}M_m\varGamma_m)}\notag\\
&\qquad\times
\int\norm{\bm{q}_1}^2\;\dd\norm{\bm{q}_1}\;\dd\varOmega\;
\dfrac{1}{2\norm{\bm{q}_1}}
\delta(M_i^2-2M_i\norm{\bm{q}_1})
\Theta(M_i-\norm{\bm{q}_1})\notag\\
&\qquad\times
\biggl[4M_i\norm{\bm{q}_1}
\pqty*{-M_i+\norm{\bm{q}_1}+\norm{\bm{q}_1}\cos\theta}
+\norm{\bm{q}_1}(1+\cos\theta)
(M_i^2-2M_i\norm{\bm{q}_1})\biggr]\notag\\
&=\dfrac{2\mathrm{i}M_i^8(M_i^2-M_m^2-\mathrm{i}M_m\varGamma_m)}
{\pi\bqty*{(M_i^2-M_m^2)^2+M_m^2\varGamma_m^2}}.
\label{eq:A.27}
\end{align}
Therefore, the imaginary part of the interference term corresponding to fig.~\ref{fig:5}~(b) is given by
\begin{equation}
\label{eq:A.28}
\Im(J_{\rm self}^{\rm (b)})
=\dfrac{1}{2\mathrm{i}}\Disc(J_{\rm self}^{\rm (b)})
=\dfrac{M_i^8}{\pi}\,\dfrac{M_i^2-M_m^2}
{(M_i^2-M_m^2)^2+M_m^2\varGamma_m^2}.
\end{equation}
Consequently, the contribution of graph (b) to the self-energy part of the CP asymmetry reads
\begin{align}
\varepsilon_{{\rm self},\alpha i}^{\gamma,{\rm (b)}}
&=-\dfrac{4}{\varGamma_i}\Im(B_{\mu}^{(5)})
\Im(J_{\rm self}^{\rm (b)}\varPhi_2)\notag\\[0.25em]
&=-\dfrac{M_i^2}{2\pi{\sum}_{\beta}\abs{\mu_{\beta i}}^2}\,
\Im\bqty*{\mu_{\alpha i}^{\ast}\mu_{\alpha m}(\mu^{\dagger}\mu)_{mi}}\,
\dfrac{(M_i^2-M_m^2)M_i^2}{(M_i^2-M_m^2)^2+M_m^2\varGamma_m^2}.
\label{eq:A.29}
\end{align}
Rewriting the Breit--Wigner form in terms of $x=M_m^2/M_i^2$, with $M_i^2-M_m^2=M_i^2(1-x)$, we obtain
\begin{align}
\tilde{f}_{S_b}(x)
&=\dfrac{(M_i^2-M_m^2)M_i^2}{(M_i^2-M_m^2)^2+M_m^2\varGamma_m^2}
\notag\\[0.25em]
&=\dfrac{1-x}{(1-x)^2+x(\varGamma_m/M_i)^2}
\simeq\dfrac{1-x}{(1-x)^2+(\varGamma_m/M_i)^2},
\label{eq:A.30}
\end{align}
where the last step holds in the quasi-degenerate limit $x\simeq 1$. This is precisely the Breit--Wigner envelope for the self-energy piece regularised by the total width $\varGamma_m$.

The results of this appendix are summarised in eqs.~(\ref{eq:A.18}) and (\ref{eq:A.30}), which are also given in eq.~(\ref{eq:3.5}) of the main text. By combining the vertex contribution calculated in app.~C.4 of Part~I with the resonant self-energy contributions computed in this appendix, all non-vanishing CP asymmetries in low-scale resonant EMLG are accounted for.

\bibliographystyle{JHEP}
\bibliography{refs}
\end{document}